\documentclass[prd,aps,floatfix,preprint,nofootinbib]{revtex4}
\usepackage{graphicx,color}
\usepackage{epstopdf}
\usepackage{amsmath,amssymb,revsymb,braket}
\usepackage{slashed}
\usepackage{subfigure}
\usepackage{epsfig}
\usepackage{float}

\def\simge{\mathrel{%
       \rlap{\raise 0.511ex \hbox{$>$}}{\lower 0.511ex \hbox{$\sim$}}}}
\def\simle{\mathrel{
       \rlap{\raise 0.511ex \hbox{$<$}}{\lower 0.511ex \hbox{$\sim$}}}}

\makeatletter
\newcommand{\figcaption}[1]{\def\@captype{figure}\caption{#1}}
\newcommand{\tblcaption}[1]{\def\@captype{table}\caption{#1}}
\newcommand{\Nf}{N_f}

\newcommand{\kq}{\kappa_q}

\newcommand{\bmq}{\bar m_q}

\newcommand{\no}{\nonumber}

\newcommand{\pd}{\partial}

\newcommand{\DU}{{\cal D}U}

\newcommand{\Dq}{\Big[\Pi_{f=1}^{N_f}{\cal D}q_f{\cal D}\bar q_f \Big]}

\newcommand{\Nsite}{N_{\rm site}}

\newcommand{\Tr}{{\rm Tr}}

\newcommand{\nn}{\nonumber}

\makeatother

\begin{document}

\title{
Topological susceptibility at high temperature on the lattice
}

\author{J.~Frison}
\author{R.~Kitano}
\author{H.~Matsufuru}
\author{S.~Mori}
\author{N.~Yamada}
\affiliation{
        High Energy Accelerator Research Organization (KEK), %
        Tsukuba 305-0801, Japan}
\affiliation{
        Graduate University for Advanced Studies (SOKENDAI), %
        Tsukuba 305-0801, Japan}

\date{\today}

\begin{abstract}
 QCD topological susceptibility at high temperature, $\chi_t(T)$,
 provides an important input for the estimate of the axion abundance in
 the present Universe.
 While the model independent determination of $\chi_t(T)$ should be
 possible from the first principles using lattice QCD, existing methods
 fail at high temperature, since not only the probability that
 non-trivial topological sectors appear in the configuration generation
 process but also the local topological fluctuations get strongly
 suppressed.
 We propose a novel method to calculate the temperature dependence of
 topological susceptibility at high temperature.
 A feasibility test is performed on a small lattice in the quenched
 approximation, and the results are compared with the prediction of the
 dilute instanton gas approximation.
 It is found that the method works well especially at very high
 temperature and the result is consistent with the instanton calculus
 down to $T\sim 2\, T_c$ within the statistical uncertainty.
\end{abstract}

\maketitle

\section{Introduction}
\label{sec:introduction}

The standard model (SM) is not invariant under P nor CP transformation.
Strangely, one of the renormalizable, CP violating terms,
$\theta\, G_{\mu\nu}\tilde G_{\mu\nu}$, exists
only with an undetectably small coefficient ($\theta\simle 10^{-10}$)
or even is missing in the SM, where $G_{\mu\nu}$ is the gluon field
strength tensor and
$\tilde G_{\mu\nu}=\epsilon_{\mu\nu\rho\sigma}G_{\rho\sigma}/2$.
This unnatural situation is referred to as strong CP problem.
Besides the solution with the vanishing up quark mass, the Peccei-Quinn
(PQ) mechanism has been known to provide an elegant
explanation~\cite{Peccei:1977hh, Peccei:1977ur,Weinberg:1977ma,
Wilczek:1982rv, Kim:1979if, Shifman:1979if,Dine:1981rt,
Zhitnitsky:1980tq}.
In the PQ mechanism, a complex scalar field with a $U(1)$ symmetry is
introduced.
After the spontaneous symmetry breaking of the $U(1)$ symmetry at very
high temperature, the radial component of the PQ field acquires the
vacuum expectation value, $f_a$, and the angular component, called axion
$a(x)$, emerges as Nambu-Goldstone (NG) boson\footnote{The axion
eventually becomes pseudo NG boson due to the axial anomaly in QCD.}.
One can rotate away the leading interaction of the axion to quarks by
the $U(1)$ chiral transformation of quark fields, and then the
coefficient of the $G\tilde{G}$ term in the SM, $\theta$, is replaced by
$\theta'=\theta+ a(x)/f_a$.
Due to the periodicity of the moduli of the axion field, the effective
potential of the axion field takes the form like
$\chi_t \cos\,\theta'$, where $\chi_t$ is the QCD
topological susceptibility.
This form leads to two important consequences.
One is that the CP conserving vacuum ({\it i.e.}
$\langle a \rangle/f_a=-\,\theta$) is automatically chosen at low
enough temperature, thus the strong CP problem is gone (PQ mechanism).
The other is that the temperature dependent axion mass is given by the
QCD topological susceptibility (and $f_a$) as
$m_a^2(T)=\chi_t(T)/f_a^2$.

The PQ mechanism is attractive because it also provides a candidate for
the dark matter of the Universe through the misalignment mechanism for
the axion generation~\cite{Preskill:1982cy, Abbott:1982af, Dine:1982ah}.
The axion abundance of the present Universe is determined by two
ingredients:
the axion mass as a function of $T$, $m_a(T)$, and
the misalignment at $T=T^*$, $\theta'|_{T\ge T^*}$, where the axion
starts coherent oscillation.

In the estimate of $\chi_t$ at finite temperature, the instanton picture
is widely adopted~\cite{'tHooft:1976fv} and predicts
$T^*\sim 6\, T_c \sim O(1)$ GeV~\cite{Pisarski:1980md, Gross:1980br,Wantz:2009it},
where $T_c\sim 150$ MeV is the (pseudo-)critical temperature for chiral
symmetry breaking in QCD.
However, the instanton calculus is based on the perturbation theory, and
hence the reliability is not very clear around $T\sim T^*\simle 1$ GeV.
Furthermore, the possibility that, in two flavor QCD, $\chi_t$ behaves
like a step function at $T=T_c$ when the quarks are sufficiently light
is argued based on reasonable assumptions \cite{Aoki:2012yj} (see also
a clarification in Ref.~\cite{Kanazawa:2015xna}).
In such a case, a significant enhancement of the axion abundance is
predicted, and even excludes the standard axion scenario if
$\theta'(T^*)=O(1)$~\cite{Kitano:2015fla}.

Numerical simulations of lattice QCD can unambiguously determine
$\chi_t(T)$ in principle.
The study of $\chi_t$ at high temperature like $\sim 2\,T_c$ or higher
began in the $SU(3)$ Yang-Mills theory~\cite{Berkowitz:2015aua,
Kitano:2015fla, Borsanyi:2015cka}.
Recently, full QCD results
were reported~\cite{Bonati:2015vqz,Petreczky:2016vrs}.
Several remarks are as follows.
First, the lattice calculations of $\chi_t(T)$ in the $SU(3)$ Yang-Mills
theory shows $\chi_t(T)\sim T^{-X}$ with $5.6\simle X \simle 7.14$, which
is compatible with $X\sim 7$ in the instanton
calculus~\cite{Pisarski:1980md, Gross:1980br, Wantz:2009it}.
Secondly, one of the full QCD calculations in
Ref.~\cite{Bonati:2015vqz} finds $X\sim 3$, which disagrees with
$X\sim 8$ in the instanton calculus, while $X\sim 8$ is reported in
Ref.~\cite{Petreczky:2016vrs}.
Thirdly, $\langle Q^2 \rangle|_T=\chi_t(T)\,V_4$ rapidly decreases with
$T$, where $V_4$ represents the four dimensional volume, and importantly
existing lattice methods fail when $\chi_t(T)V_4\ll 1$
\footnote{An interesting proposal to avoid this difficulty is found in
Ref.~\cite{Laio:2015era}.}.
Since $\chi_t(T) V_4\ll 1$ is realized above a certain temperature
before reaching $T^*$, the axion abundance becomes uncertain.
Thus, methods overcoming this difficulty are desired.

To be specific, in Ref.~\cite{Kitano:2015fla}, where $\chi_t(T)$ is
calculated on $16^3\times 4$ lattices in the quenched approximation with
one of the standard methods counting the fermionic zero modes,
$\chi_t(T)V_4$ is estimated to be 0.35, 0.09, 0.03 at $T=1.34\,T_c$,
$1.5\,T_c$, $1.75\,T_c$, respectively, and no reliable estimate is given
above $2\,T_c$.

In the present paper, we propose a novel method to determine the
temperature dependence of $\chi_t(T)$ at high temperature.
The method involves estimating the difference of the gauge action and
the chiral condensate between two different topological sectors.
In order to see how well the proposed method works, we perform a test in
pure Yang-Mills theory on a small lattice.
The results are found to be reasonably consistent with the instanton
calculus above $T\sim 2\,T_c$.

The paper is organized as follows.
In sec.~\ref{sec:instanton}, some ingredients of the instanton
calculus relevant to subsequent sections are briefly reviewed.
After the method is described in sec.~\ref{sec:method}, the numerical
test is presented in sec.~\ref{sec:test}.
Summary and outlook are given in sec.~\ref{sec:summary}.
To supplement the main text, three sections are put in appendix, which
include the explicit form of lattice Dirac operators, the discussion
of the $\chi_tV_4\gg 1$ case in this framework, and the analysis of the
hopping parameter expansion.

\section{Instanton calculation}
\label{sec:instanton}

For later use, the calculation of the topological susceptibility in the
dilute instanton-gas approximation (DIGA)~\cite{Gross:1980br} is briefly
reviewed.
In general, the gauge action is bounded by
\begin{eqnarray}
 S_g
&=& \int d^4x \frac{1}{2}\Tr\,\left[ G_{\mu\nu}G_{\mu\nu} \right]
\ge \int d^4 x \frac{1}{2} \left|\Tr\, G_{\mu\nu}\tilde{G}_{\mu\nu}\right|
\ ,
\end{eqnarray}
where the equality holds when $G_{\mu\nu}=\pm\tilde G_{\mu\nu}$.
This self-duality relation is realized in the BPST instanton solution
\cite{Belavin:1975fg}.
In instanton calculus, the BPST instanton is taken as the classical
background, and the effects of quantum or thermal fluctuations around it
is incorporated perturbatively.

Using the partition functions with the topological charge $Q=0$ and $1$
\footnote{For the definition of $Z_Q$, see eq.~(\ref{eq:zq}).},
the instanton density $n(\rho,T)$ for the one-instanton sector is
defined by
\begin{eqnarray}
 \frac{Z_{Q=1}}{Z_{Q=0}} = \int \frac{d^4z\,d\rho}{\rho^5}\,n(\rho,T),
\\
 n(\rho,T) = n_G(\rho)\,n_F(m_f\rho)\,n_T(\pi\rho T)
\ ,
\end{eqnarray}
where $z$ and $\rho$ are the position and the size of the instanton,
respectively, and $n(\rho,T)$ is factorized into the gauge ($n_G$),
the fermionic part ($n_F$) and the finite temperature effect ($n_T$).

After the explicit calculation of $Z_{Q=1}$
\cite{'tHooft:1976fv,Bernard:1979qt}, the gauge contribution to the
instanton density is found to be
\begin{eqnarray}
    n_G(\rho) 
&=& C_I\ (\mu\rho)^{\beta_0'} \left(\frac{8\pi^2}{g^2(\mu)}\right)^{2N}
    e^{-8\pi^2/g^2(\mu)},
\\
    C_I
&=& \frac{1}{4^N}\frac{2\,e^{5/6}}{\pi^2 (N-1)! (N-2)!}
    e^{-2N\alpha + \frac{N}{6}}
\ ,
\label{eq:c1}
\end{eqnarray}
where $N$ is the number of colors and the renormalization scale $\mu$ is
introduced.
Other undefined constants are shortly given.
The constant $N/6$ in eq.~(\ref{eq:c1}) does not appear in the
Pauli-Villars regularization and appears in the $\overline{\rm MS}$
scheme~\cite{Luscher:1982wf}.

In the presence of $N_f$ flavors of quarks with the mass $m_f$
\cite{'tHooft:1976fv,Carlitz:1978yj,Novikov:1983gd,Kwon:2000kf,Dunne:2005te},
the fermionic contribution is given, using the Pad\'{e} approximation
\cite{Kwon:2000kf,Dunne:2005te}, by
\begin{eqnarray}
  n_F(m_f\rho)=
   \prod_{f=1}^{N_f}(m_f\rho)^{\frac{2}{3}}
   \exp\left(2\frac{   \frac{1}{6}\log(m_f\rho) + \alpha
	             - (3\alpha+c) (m_f\rho)^2 + a_1 (m_f\rho)^4
		     - a_2 (m_f\rho)^6}
		   { 1 - 3 (m_f\rho)^2 + b_1 (m_f\rho)^4
		       + b_2 (m_f\rho)^6 + b_3 (m_f\rho)^8}
       \right) 
\label{eq:dunne_result}
\ .
 \end{eqnarray}
The constants involved in the above equations are
\begin{eqnarray}
&&  \beta_0'
 =  \beta_0+\left(\beta_1-4\beta_0 N+\gamma_0 N_f\right)
    \frac{g^2(\mu)}{16\pi^2},\\
&&  \beta_0
 =  \frac{11}{3}N - \frac{2}{3} N_f,\ \
    \beta_1
 =   \frac{34}{3}N^2
   - \left(\frac{13}{3}N - \frac{1}{N}\right)N_f,\ \
    \gamma_0
 =  3\frac{N^2-1}{N}
\ ,
\\&&     \alpha
\equiv \alpha\left(\frac{1}{2}\right)\simeq 0.145873,\quad c
  \equiv \frac{1}{2}(\ln2-\gamma)\simeq 0.05797\ ,
\\&&
  a_1 = -13.4138,\quad a_2 = 2.64587,
\\&&
  b_1 = 25\left(\frac{592955}{21609}a_2 + \frac{255}{49}a_1 + 9\alpha +
  3\beta_0'\right),\quad b_2 = -75\left(\frac{85}{49} a_2 +
  a_1\right),\quad b_3 = 75 a_2
\ .
\end{eqnarray}

In the finite temperature QCD, the explicit form of the instanton on
$S^1\times R^3$ is known as the HS caloron~\cite{Harrington:1978ve}.
While the $\rho$ integral diverges at zero temperature, it becomes
finite at finite temperature~\cite{Gross:1980br} since the Debye
screening exponentially suppresses the large size instanton.
This effect is embedded in $n_T$, which is known to be
\begin{eqnarray}
    n_T(\lambda)
&=& \exp\left[ - \frac{1}{3}(2N+N_f)\lambda^2
	       - 12 A(\lambda) \left(1 + \frac{1}{6}(N-N_f)\right)
	\right],\nn\\
   A(\lambda) &=&
      - \frac{1}{12}\log\left(1+\frac{\lambda^2}{3}\right)
      + c_1 \left(\frac{1}{1+c_2 \lambda^{-\frac{3}{2}}}\right)^{8}
\ ,
\end{eqnarray}
where $\lambda=\pi\rho T$, $c_1 = 0.01289764$, and $c_2 = 0.15858$.

Collecting the above expressions, the DIGA predicts
the topological susceptibility $\chi_t(T)$ at finite temperature to be
\begin{eqnarray}
        \chi_t(T)V_4 
\approx \frac{Z_{Q=1}+Z_{Q=-1}}{Z_{Q=0}}
 =      2 \int\!\! d^4z\! \int^\infty_0 \frac{d\rho}{\rho^5}\ n(\rho,T)
\ .
\label{eq:chit-diga}
\end{eqnarray}
Later, $d\ln\chi_t(T)V_4/d\ln T$ in the DIGA is numerically estimated to
compare with the lattice result, where the running coupling
is calculated with the four loop $\beta$ function.
Focusing on the temperature dependence in the high temperature limit
where the DIGA is reliable, it follows from eq.~(\ref{eq:chit-diga})
that
 \begin{eqnarray}
    \lim_{T\to \infty} \chi_t(T)
&\propto& 
    \int_0^{\rho_{\rm cut}}\! d\rho\
    \rho^{\beta_0'-5}\, (\pi\rho T)^{2+\frac{1}{3}(N-N_f)}
\no\\
&=&  \frac{1}{\beta_0' + \frac{1}{3}(N-N_f)-2}\,
    \rho_{\rm cut}^{\beta_0' + \frac{1}{3}(N-N_f)-2}
    (\pi T)^{2+\frac{1}{3}(N-N_f)},\nn\\
&\propto&
    T^{4-\beta_0'},
\end{eqnarray}
where $\rho_{\rm cut}=\sqrt{3}/(\sqrt{2N+N_f}\,\pi T)$.

\section{Method}
\label{sec:method}

We begin with clarifying our conventions and notations.
The QCD partition function on the lattice with a finite $\theta$ and in
the specific topological sector can be written as
\begin{eqnarray}
    Z_\theta(\beta,\bar m_q)
&=& \sum_{Q=-\infty}^{+\infty} Z_Q(\beta,\bar m_q)\,
    e^{-i\theta Q}\
 =  e^{-V_4\,E(\theta)}
\label{eq:ztheta}
\ ,\\
    Z_Q(\beta,\bar m_q)
&=& \int_{\in Q}\!\!\DU\Dq\,
    e^{- S_g(\beta) - S_q(\bar m_q)}
\no\\
&=& \frac{1}{2\pi}\int\!\! d\theta\, Z_\theta(\beta,\bmq) e^{i\theta Q}
 =  \frac{1}{2\pi}\int\!\! d\theta\, e^{i\theta Q-V_4\,E(\theta,T)}
\label{eq:zq}
\ ,
\end{eqnarray}
respectively, where $\in Q$ denotes that the integral is restricted to
the configurations with a topological charge $Q$, and$E(\theta)$
($-\pi< \theta \le \pi$) is the internal energy density.
$E(\theta)$ is periodic in $\theta$ with periodicity of $2\pi$ and
symmetric under the change of the sign of $\theta$, {\it i.e}
$E(\theta)=E(-\theta)=E(2\pi+\theta)$.

The lattice gauge action $S_g$ is given by
\begin{eqnarray}
    S_g
&=& 6\,\beta\,\Nsite\,\big\{ (c_0+2c_1) - P\big\}
\label{eq:gauge-action}
\ ,
\end{eqnarray}
where $\beta=6/g^2$ represents the lattice gauge coupling.
The action density $P$ is given by
\begin{eqnarray}
     P
&=&  c_0\, W_P + 2\, c_1\, W_R\
\label{eq:action-density}
\ ,
\end{eqnarray}
where $W_P$ and $W_R$ denote the $1\times 1$ plaquette and $1\times 2$
rectangle averaged over four-dimensional lattice sites, respectively.
$c_0$ and $c_1$ satisfying $c_0=1-8 c_1$ are the improvement
coefficients for the lattice gauge action.
The total number of lattice sites is
$\Nsite= N_S^3 \times N_T = V_4(\beta)/a^4(\beta)$, where $N_S$ and
$N_T$ represent the number of lattice sites in the spatial and time
directions, respectively.
For fixed $N_S$ and $N_T$, the physical, four dimensional volume
$V_4(\beta)$ and the lattice spacing $a(\beta)$ depend only on $\beta$.
The temperature of the system in the physical unit is given by
\begin{eqnarray}
    T(\beta,N_T)
&=& \frac{1}{a(\beta)N_T}
\label{eq:temperature}
\ ,
\end{eqnarray}
and hence it can be changed by adjusting either the temporal size $N_T$ or
the lattice bare coupling $g^2=\beta/6$ \footnote{Thus, we take the mass
independent scale setting prescription, where the lattice spacing $a$
does not depend on the quark mass.}.
$S_q$ is the lattice quark action, which we do not specify here.
In the following, we consider $N_f$ flavors of quarks with a degenerate
mass $m_q$ ($\bmq=a(\beta)m_q$ in the lattice unit) for simplicity.
The extension to non-degenerate cases is straightforward.

Assuming $\theta=0$, the expectation value of an operator $O$ is
expressed as
\begin{eqnarray}
    \langle O \rangle_{\beta,\bmq}^{\theta=0}
&=& \frac{1}{Z_{\theta=0}(\beta,\bmq)}
    \sum_{Q=-\infty}^{+\infty} Z_Q(\beta,\bmq)
    \langle O \rangle_{\beta,\bmq}^{(Q)}
\end{eqnarray}
where we have defined
\begin{eqnarray}
    \langle O \rangle_{\beta,\bmq}^{(Q)}
&=& \frac{1}{Z_Q(\beta,\bmq)}
    \int_{\in Q}\!\!\DU\Dq\,
    e^{- S_g(\beta)-S_q(\bmq)}\, O
\label{eq:expect-value-with-Q}
\ .
\end{eqnarray}
Thus, the topological susceptibility times four dimensional volume
is written as
\begin{eqnarray}
    \chi_t(\beta,\bmq)V_4(\beta)
&=& \langle Q^2 \rangle_{\beta,\bmq}^{\theta=0}
 =  \frac{1}{Z(\beta,\bmq)}
    \sum_{Q=-\infty}^{+\infty} Z_Q(\beta,\bmq)\, Q^2
\no\\
&=& \frac{      Z_1(\beta,\bar m_q) +     Z_{-1}(\beta,\bar m_q)
          + 4\, Z_2(\beta,\bar m_q) + 4\, Z_{-2}(\beta,\bar m_q)
	  + \cdots}
         {\sum_{Q=-\infty}^{+\infty} Z_Q(\beta,\bar m_q)}
\label{eq:chitv-0}
\ .
\end{eqnarray}

The simplest method to calculate $\chi_t$ is to generate an ensemble on
the lattice and look at the distribution of $Q$.
As is seen, for example, in Fig.~1 of Ref.~\cite{Kitano:2015fla}, an
update algorithm employed there only generate configurations with $Q=0$
or $\pm 1$ at some high temperature~\footnote{Note that, even in such a
case, the resulting value of $\chi_t$ turns out to be consistent with
more extensive lattice simulations such as
Refs.~\cite{Berkowitz:2015aua,Borsanyi:2015cka}.}.
Since those with $Q=0$ dominates the other, $Z_0\gg Z_{\pm 1}$ should
hold, and it follows from eq.~(\ref{eq:chitv-0})
\begin{eqnarray}
    \frac{Z_{\pm 1}(\beta,\bar m_q)}{Z_0(\beta,\bar m_q)}
&\approx&
    \frac{w(\beta,\bmq)}
         {2}
\label{eq:fact}
\ ,
\end{eqnarray}
where we have defined $w(\beta,\bmq) = \chi_t V_4$.

So far, the partition function, $Z_Q$, has been written as a function of
$\beta$ and $\bmq$, but an arbitrary pair of arguments can be chosen as
long as they fix the QCD coupling and the quark masses.
In the following, we consider $(T,\ m_q)$ and $(w=\chi_tV_4,\ m_q)$ as a
pair of arguments, and fix $m_q$ to the physical quark mass as function
of $T$ or $w$.
In this case, $Z_Q$ can be viewed as the function of $T$ or $w$.
Furthermore, the numbers of lattice sites in the spatial and the
temporal directions, {\it i.e.} $N_T$ and $N_S$, are also fixed.

Consider the derivative of the ratio of the partition functions of
different topological sectors with respect to temperature with $m_q$ and
$\Nsite$ fixed, $d\ln (Z_{Q_2}/Z_{Q_1})/d\ln T\big|_{\Nsite}$.
Using the
chain rule, we rewrite it as
\begin{eqnarray}
    \frac{d\ln \frac{Z_{Q_2}(T)}{Z_{Q_1}(T)}}
         {d\ln T}\bigg|_{\Nsite}
&=& \frac{d\ln w(T)}
         {d\ln T}\bigg|_{\Nsite}
    \frac{d\ln \frac{Z_{Q_2}(w)}{Z_{Q_1}(w)}}
         {d\ln w}\bigg|_{\Nsite}
\ .
\end{eqnarray}
Then, the $T$ dependence of $w$ is expressed as
\begin{eqnarray}
    \frac{d\ln w(T)}{d \ln T}\bigg|_{\Nsite}
&=& \frac{d\ln\frac{Z_{Q_2}(T)}{Z_{Q_1}(T)}}{d \ln T}\bigg|_{\Nsite} \times
    \left(\frac{d\ln\frac{Z_{Q_2}(w)}{Z_{Q_1}(w)}}{d \ln w}\bigg|_{\Nsite}\right)^{-1}
\ .
\label{eq:dzdt_vs_dwdt}
\end{eqnarray}
In the following, the symbol $\big|_{\Nsite}$ is omitted for
simplification.
How to estimate each of two factors in the r.h.s. is described below.

The first factor, $d\ln (Z_{Q_2}/Z_{Q_1})/d \ln T$, can be calculated
using lattice numerical simulations as follows, where the temperature,
defined in eq.~(\ref{eq:temperature}), is varied by changing $\beta$
while $N_T$ is fixed.
The differentiation of $Z_Q$ with respect to $T$ is then given by
\begin{eqnarray}
    \frac{d\ln Z_Q(T)}{d \ln T}
&=& \bigg(
     \frac{d\beta}{d \ln T}
     \frac{\pd}{\pd \beta}
   + \frac{d \ln \bar m_q}{d \ln T}
     \frac{\pd}{\pd \ln \bar m_q}
    \bigg)
    \ln Z_Q(\beta, \bmq)\ .
\label{eq:dynamical-case}
\end{eqnarray}
The $\beta$ derivative term in eq.~(\ref{eq:dynamical-case}) is found to
be
\begin{eqnarray}
      \frac{d\beta}{d \ln T}\frac{\pd\ln Z_Q(\beta,\bmq)}{\pd \beta}
&=& - \frac{\beta\,\beta_g}{6}\,
      \langle S_g \rangle_{\beta,\bmq}^{(Q)}
\label{eq:dbdSg}
\ ,
\end{eqnarray}
where we have used eq.~(\ref{eq:gauge-action}),
eq.~(\ref{eq:temperature}) and the $\beta$ function for the QCD coupling
\begin{eqnarray}
    \beta_g
&=& \frac{d g^2}{d \ln a}
 =  2 g \frac{d g}{d \ln a}
\ ,
\end{eqnarray}
In perturbation theory, the first two coefficients of $\beta_g$ are
given by
\begin{eqnarray}
    \beta_g
&=& 2\, b_0\, g^4 + 2\, b_1\, g^6 + O(g^8)
\ ,\\
    b_0
&=& \frac{11-\frac{2}{3}N_f}{(4\pi)^2},\ \
    b_1
 =  \frac{102-\frac{38}{3}N_f}{(4\pi)^4}\
.
\end{eqnarray}
For our purpose, $\beta_g$ has to be numerically determined as the
temperature considered here is of $O(T_c)$.

The term including the mass derivative in eq.~(\ref{eq:dynamical-case})
are estimated as follows.
The first factor is found to be
\begin{eqnarray}
    \frac{d\ln \bar m_q}{d \ln T}
&=& \frac{d\ln a}{d \ln T}\frac{d \ln \bar m_q}{d\ln a}
 =  - \bigg( 1 + \frac{d \ln m_q}{d\ln a} \bigg)
\label{eq:anomalousd-mass}
\ ,
\end{eqnarray}
which is related to the anomalous dimension of the quark mass.
The second factor is calculated to be
\begin{eqnarray}
    \frac{\pd\ln Z_Q(\beta,\bar m_q)}{\pd \ln \bar m_q}
&=& - N_f\,\bmq
      \langle s_{\bar qq} \rangle^{(Q)}_{\beta,\bar m_q}
\label{eq:dbmq-Sq}
\ ,
\end{eqnarray}
where the explicit form of the scalar density operator,
$s_{\bar qq}$, requires specifying the quark action, $S_q$.
For example, it is given by
\begin{eqnarray}
    s_{\bar qq}
&=& \sum_{x} \bar {q}_x\,{q}_x
\label{eq:qqbar-def-wil}
\ ,
\end{eqnarray}
for the Wilson fermion, and
\begin{eqnarray}
    s_{\bar qq}
&=& \sum_{x,y}
      \bar {q}_x 
      \bigg(\delta_{x,y} - \frac{1}{2M_0}D^{\rm ov}_{x,y}(0)\bigg)
      {q}_y
\label{eq:qqbar-def}
\ ,
\end{eqnarray}
for the overlap fermion.
For details, see appendix~\ref{sec:dynamical}.

Gathering eqs.~(\ref{eq:dbdSg}), (\ref{eq:anomalousd-mass}) and
(\ref{eq:dbmq-Sq}), eq.~(\ref{eq:dynamical-case}) becomes
\begin{eqnarray}
    \frac{d\ln Z_Q(T)}{d \ln T}
&=& - \frac{\beta\,\beta_g}{6}\,
      \langle S_g \rangle_{\beta,\bmq}^{(Q)}
    + N_f\,\bigg( 1 + \frac{d \ln m_q}{d\ln a} \bigg)
      \bmq \langle s_{\bar qq} \rangle^{(Q)}_{\beta,\bar m_q}
\label{eq:dynamical-case-2}
\ .
\end{eqnarray}
Taking the difference of eq.~(\ref{eq:dynamical-case-2}) for $Q_2$ and
$Q_1$, we obtain
\begin{eqnarray}
    \frac{d\ln \frac{Z_{Q_2}}{Z_{Q_1}}}{d \ln T}
&=&
     \frac{\beta^2\,\beta_g}{6}\Delta S_g^{(Q_2,Q_1)}(\beta,\bar m_q)
   + N_f\,\bigg( 1 + \frac{d \ln m_q}{d\ln a} \bigg)\,
     \bmq
     \bigg(
        \langle s_{\bar qq} \rangle^{(1)}_{\beta,\bar m_q}
      - \langle s_{\bar qq} \rangle^{(0)}_{\beta,\bar m_q}
     \bigg)
\label{eq:tdep-of-chit-prep}
\ ,
\end{eqnarray}
where we have defined
\begin{eqnarray}
      \Delta S_g^{(Q_2,Q_1)}(\beta,\bar m_q)
&=& - \frac{1}{\beta}
      \bigg(
          \langle S_g \rangle^{(Q_2)}_{\beta,\bar m_q}
        - \langle S_g\rangle^{(Q_1)}_{\beta,\bar m_q}
      \bigg)
\label{eq:deltaSg}\ ,
\end{eqnarray}
for later use.
From eq.~(\ref{eq:tdep-of-chit-prep}), it turns out that the differences
of the gauge action and the chiral condensate between two topological
sectors are required to determines the temperature dependence of
$Z_{Q_2}/Z_{Q_1}$.

Next, we turn to the second factor in eq.~(\ref{eq:dzdt_vs_dwdt}),
$d\ln(Z_{Q_2}/Z_{Q_1})/d \ln w$.
In the following, the arguments of the partition functions are omitted
for the sake of simplicity.
When $w\gg 1$, existing lattice methods should work well, and our method
is not more efficient than those.
However, since it is still instructive to analyze the $w\gg 1$ case
within the new framework, several remarks are described in the
appendix~\ref{sec:wgg1}.
Hereafter, we focus on the $w\ll 1$ case.
Note that, although $w\ll 1$, we assume that the spatial volume is still
larger than the typical length scale of the system ($\sim 1/T$).

Now assume that $Z_Q$ can be expanded in terms of $w$ as
\begin{eqnarray}
&& 
 Z_Q = a_Q w^{n_Q} + O(w^{n_Q+1})\ ,
 \label{eq:wscaling}
\end{eqnarray}
with an unknown coefficient $a_Q$.
While $n_Q$ for arbitrary $Q$ is not known, previous numerical
simulations tell that there is a temperature region where
$Z_{\pm 1}/Z_0\approx w/2$ holds [eq.~(\ref{eq:fact})], indicating
$n_{\pm 1}-n_0=1$.
With the assumption eq.~(\ref{eq:wscaling}), it follows that
\begin{eqnarray}
    \left(\frac{d\ln\frac{Z_{Q_2}}{Z_{Q_1}}}{d \ln w}\right)
&=& n_{Q_2} - n_{Q_1} + O(w)
\label{eq:dzdw-small-w}
\end{eqnarray}

Using eqs.~(\ref{eq:tdep-of-chit-prep}) and (\ref{eq:dzdw-small-w}) and
recalling $d\ln V_4/d\ln T= - 4$, eq.~(\ref{eq:dzdt_vs_dwdt}) is
rewritten as
\begin{eqnarray}
    \frac{d\ln \chi_t(T)}{d \ln T}
&=&
  \Bigg[
     \frac{\beta^2\,\beta_g}{6}\Delta S_g^{(Q_2,Q_1)}(\beta,\bar m_q)
   + N_f\,\bigg( 1 + \frac{d \ln m_q}{d\ln a} \bigg)\,
     \bmq
     \bigg(
        \langle s_{\bar qq} \rangle^{(1)}_{\beta,\bar m_q}
      - \langle s_{\bar qq} \rangle^{(0)}_{\beta,\bar m_q}
     \bigg)
  \Bigg]
\no\\&&\hspace{0ex}
  \times \frac{1}{n_{Q_2}-n_{Q_1}}
  + 4 + O(w)
\label{eq:tdep-of-chit}
\ .
\end{eqnarray}
If the boundary condition for this differential equation is provided, we
can determine the absolute value of $\chi_t(T)$.

It should be noted that the l.h.s. of eq.~(\ref{eq:tdep-of-chit}) is
independent of the choice of $Q_1$ and $Q_2$ up to $O(w)$.
By equating the r.h.s. of eq.~(\ref{eq:dzdt_vs_dwdt}) for
different pairs of $Q$, we can numerically determine the ratio
$(d\ln(Z_{Q_1}/Z_{Q_2})/d\ln w)/(d\ln (Z_{Q_3}/Z_{Q_4})/d\ln w)$ by
\begin{eqnarray}
    R^{(Q_1,Q_2,Q_3,Q_4)}(\beta)
&=& \frac{d\ln\frac{Z_{Q_1}}{Z_{Q_2}}}{d \ln T} \times
    \left(\frac{d\ln\frac{Z_{Q_3}}{Z_{Q_4}}}{d \ln T} \right)^{-1}
 =  \frac{d\ln\frac{Z_{Q_1}}{Z_{Q_2}}}{d\ln w}\times
    \left(\frac{d\ln\frac{Z_{Q_3}}{Z_{Q_4}}}{d\ln w}\right)^{-1}
\label{eq:ratio}
\ ,
\end{eqnarray}
independently of the size of $w$.
Then, the assumption eq.~(\ref{eq:wscaling}) gives 
\begin{eqnarray}
    R^{(Q_1,Q_2,Q_3,Q_4)}(\beta)
&=& \frac{n_{Q_1}-n_{Q_2}}{n_{Q_3}-n_{Q_4}} + O(w)
\label{eq:ratio-2}
\ .
\end{eqnarray}
Especially, when $Q_2=Q_4=0$ and $Q_3=1$,
$R^{(Q_1,0,1,0)}(\beta)=n_{Q_1}-n_0$.
Thus, measuring $R^{(Q,0,1,0)}(\beta)$ with various $Q$ enables us to
investigate the leading power of $Z_Q/Z_0$, {\it i.e.} $n_Q-n_0$.
On the other hand, when $w\gg 1$, the behavior of
$R^{(Q_1,Q_2,Q_3,Q_4)}(\beta)$ becomes \footnote{See
eq.~(\ref{eq:wgg1}) in the appendix~\ref{sec:wgg1} for more details.}
\begin{eqnarray}
    R^{(Q_1,Q_2,Q_3,Q_4)}(\beta)
&\propto& \frac{Q_1^2-Q_2^2}{Q_3^2-Q_4^2} + O(1/w)
\label{eq:ratio-3}
\ .
\end{eqnarray}
In this case, calculating $R^{(Q_1,Q_2,Q_3,Q_4)}(\beta)$ may serve to
check whether $w\gg 1$ indeed holds.

Here let us comment on our method.
If one could calculate the right hand side of
eq.~(\ref{eq:tdep-of-chit-prep}) over a wide range of $T$,
$Z_{Q_2}/Z_{Q_1}$ can be obtained by the numerical integration with an
suitable input.
By repeating this procedure for arbitrary pairs of ($Q_1,\ Q_2$) and
substituting $Z_{Q_2}/Z_{Q_1}$ thus obtained into
eq.~(\ref{eq:chitv-0}), one can determine $\chi_t(T)$ over a wide range
of $T$ without any assumptions, in principle.
If that is possible, the most of above arguments are unnecessary.
However, as we will show soon, it turns out that the numerical accuracy
is rather limited and the above naive procedure does not work well.

In this work, we instead focus on $d\ln\chi_t(T)/d\ln T$ in the
temperature region, where $\chi_t(T)V_4\approx 2\, Z_{\pm 1}/Z_0$ is
valid, because this quantity still provides useful information.
For example, the leading powers of $w$ in $Z_{Q_2}/Z_0$, {\it i.e.}
$n_{Q_2}-n_0$, extracted through eq.~(\ref{eq:ratio-2}) for various
$Q_2$ ($Q_1$ is fixed to zero for simplicity) can be used to identify
the $\theta$ dependence of the energy density\footnote{The general form
of it is given in eq.~(\ref{eq:fn4}).}.
Furthermore, once an integer value of $n_{Q_2}-n_0$ was determined,
$d \ln (Z_{Q_2}/Z_0)/d \ln T$ provides an independent determination of
$d \ln (Z_{\pm 1}/Z_0)/d \ln T$ through eq.~(\ref{eq:tdep-of-chit})
with $n_{Q_1}=n_0$ as we will explicitly show in the next section.

\subsection{high temperature limit}

It is instructive to see the high temperature limit of
eq.~(\ref{eq:tdep-of-chit}).
In this limit, the gauge action in each topological sector is
expected to realize the BPST instanton solution, at least in the
continuum theory, {\it i.e.}
$\langle S_g \rangle^{(Q)}_{\beta,\bmq}\to \frac{8\pi^2}{g^2}|Q|$.
Thus, $\langle S_g \rangle^{(Q)}_{\beta,\bmq}/\beta$ has a finite
value in the high $T$ limit,
\begin{eqnarray}
    \lim_{T\to \infty}
    \frac{1}{\beta}
    \langle\, S_g\,\rangle^{(Q)}_{\beta,\bmq}
&=& \frac{4\pi^2}{3}|Q|
\ .
\end{eqnarray}
Using the perturbative expression for $\beta_g$ and keeping only the
leading order contribution, $\beta^2\beta_g$ takes
\begin{eqnarray}
      \lim_{T\to \infty} \beta^2 \beta_g 
&=& \frac{11-\frac{2}{3}N_f}{(4\pi)^2}\times 72
\ .
\end{eqnarray}
Collecting the above yields
\begin{eqnarray}
    \lim_{T\to \infty}
    \frac{d\ln \chi_t(T)}{d \ln T}
&=& \frac{1}{n_{Q_2}-n_{Q_1}}
    \Bigg[\
      (|Q_2|-|Q_1|)\left(\frac{2}{3}N_f - 11 \right)
\no\\&&\hspace{12ex}
    + N_f\,
      \lim_{T\to\infty}
      \bmq \,
      \big(
         \langle s_{\bar qq} \rangle^{(Q_2)}_{\beta,\bar m_q}
       - \langle s_{\bar qq} \rangle^{(Q_1)}_{\beta,\bar m_q}
      \big)
    \Bigg] + 4
\ ,
\label{eq:lattice-expression}
\end{eqnarray}
where the $O(w)$ contribution is omitted.

With $N_f=0$, the r.h.s. of eq.~(\ref{eq:lattice-expression}) gives
$-11\times (|Q_2|-|Q_1|)/(n_{Q_2}-n_{Q_1})+4$.
In this case, instanton calculus predicts $\chi_t\sim T^{-7}$, which
is reproduced when $n_{Q}=|Q|$.
The instanton calculus for $N_f=0$ should also be reproduced in the
heavy quark limit, in which the heavy quarks will be decoupled from the
theory and hence the $\beta$-function is reduced to the one for $N_f=0$.
By imposing that the heavy quark limit of
Eq.~(\ref{eq:lattice-expression}) yields $\chi_t\sim T^{-7}$,
\begin{eqnarray}
    \lim_{\bmq\to \infty}\lim_{T\to\infty}
    \bmq \,
    \big(
       \langle s_{\bar qq} \rangle^{(Q_2)}_{\beta,\bar m_q}
     - \langle s_{\bar qq} \rangle^{(Q_1)}_{\beta,\bar m_q}
    \big)
&=& O(1/\bmq)
\ ,
\end{eqnarray}
is obtained.
The vicinity of the heavy quark limit can be analyzed by applying the
hopping parameter expansion, which is described in the appendix
\ref{sec:hpe}.

When $N_f=3$, the instanton calculus predicts $\chi_t \sim T^{-8}$,
which indicates
\begin{eqnarray}
    \lim_{T\to\infty}
    \bmq \,
    \big(
       \langle s_{\bar qq} \rangle^{(Q_2)}_{\beta,\bar m_q}
     - \langle s_{\bar qq} \rangle^{(Q_1)}_{\beta,\bar m_q}
    \big)
&=& - \big(n_{Q_2}-n_{Q_1}\big) + O(\bmq)
\ .
\end{eqnarray}
This coincides with the contributions from the fermion zero modes,
$-(|Q_2|-|Q_1|)$, when $n_Q = |Q|$.

\section{test in the quenched approximation}
\label{sec:test}

\subsection{lattice setup}
\label{subsec:lattice-setup}

In order to see how well the method described in the previous section
works, we perform a test in the quenched approximation.
The configurations are generated using the renormalization group
improved Iwasaki gauge action, {\it i.e.} $c_1=-0.331$.
The lattice volume is fixed to $16^3\times 4$ in this feasibility test
except one simulation, in which the calculation is repeated on
$24^3\times 4$ lattice to see the volume dependence.
However, the number of configurations required for a fixed statistical
error grows as $\Nsite$, and our limited computational resources did not
allow us to investigate the size dependence in detail.

We use the index theorem in defining the topological charge,
$Q=$ Index$[D_{\rm ov}]$, where $D_{\rm ov}$ is the overlap Dirac
operator shown in (\ref{eq:Dov}).
Since the configurations in a fixed topological sector is needed, we
insert the topology fixing (TF) term,
\begin{eqnarray}
 \frac{\det\left[H_W(-M_0)^2\right]}
      {\det\left[H_W(-M_0)^2+\mu^2 \right]}\ ,
 \label{eq:topfix}
\end{eqnarray}
into the path integral~\cite{Fukaya:2006vs} to fix $Q$ during the update
process.
The explicit form of the Hermitian Wilson Dirac operator, $H_W$, is
found in eq.~(\ref{eq:Hw}).
Due to this term, the appearance of the eigenvalues of $H_W$ smaller
than $\mu$, $|\lambda_{H_W}|\simle\mu$, is suppressed, and so is the
topology change.
In this work, $\mu=0.2$ and $M_0=1.6$ are used.
The standard hybrid Monte Carlo (HMC) method is applied in the
configuration generation.
The step size in the molecular dynamics procedure is tuned to realize
the acceptance ratio of 75\% to 90 \%.

In the preparation step, we first generate configurations at around
$T_c$ without the TF term to sample the configurations with various $Q$
values.
Then, the TF term is turned on, and $\beta$ is changed to a desired
value.
The topological charge of configurations thus generated is monitored by
calculating the index of the overlap Dirac operator [see
eq.~(\ref{eq:Dov})] with the same value of $M_0$ as that in the TF term,
and we checked that no transition to a different $Q$ sector occurs
within the configurations used in the analysis except in the $Q=-2$
sector on $24^3\times 4$ lattice, where $Q=-2$ is changed to $-1$ after
1,310 trajectories.

In the following plots, we present the statistical error only, which is
estimated by the standard single elimination jackknife method with the
bin size of 50 trajectories.
Increasing the bin size by a factor two only changes the size of
uncertainty by a few \%.

The theory with the TF term (\ref{eq:topfix}) is not rigorously
equivalent to the quenched QCD, because the TF term (\ref{eq:topfix})
would break $Z_3$ symmetry.
Thus, strictly speaking, the action with the TF term may not allow us to
study the phase transition of the quenched QCD.
Thus, our study focuses on the temperature region like $T\ge 2\,T_c$.

\begin{figure}[tbh]
 \begin{center}
  \begin{tabular}{rl}
  \includegraphics*[width=0.5 \textwidth,clip=true]
  {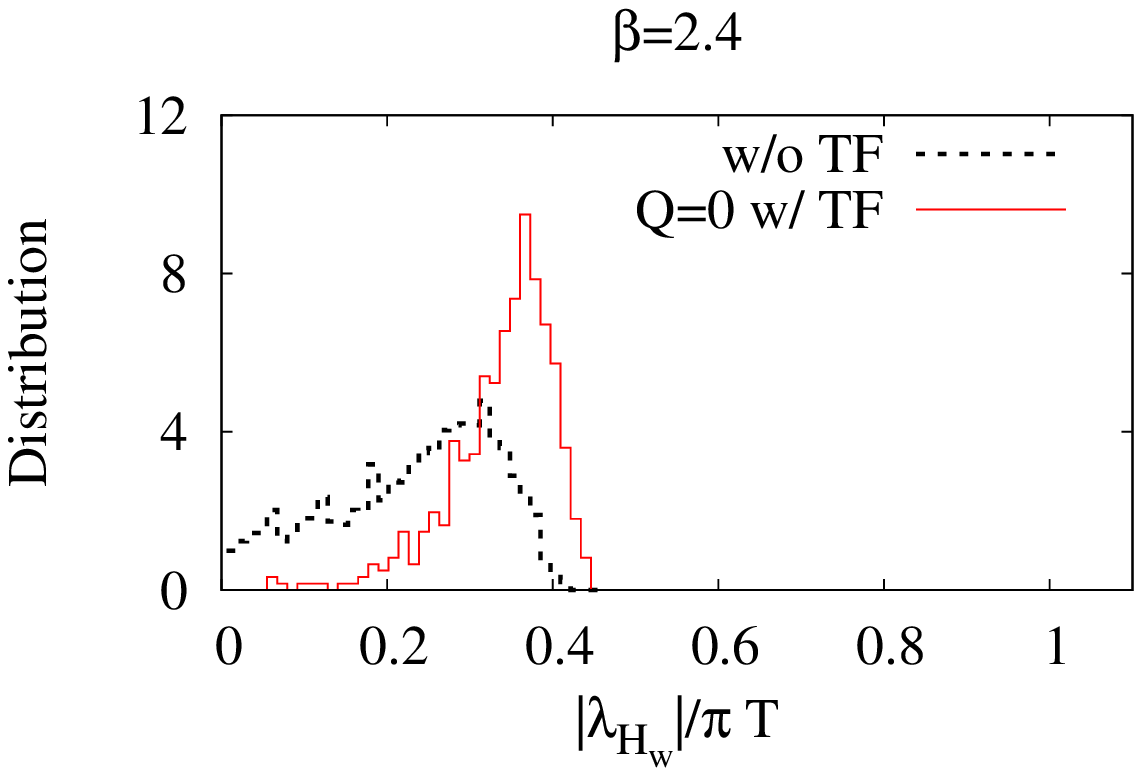}&\hspace{-4ex}
  \includegraphics*[width=0.5 \textwidth,clip=true]
  {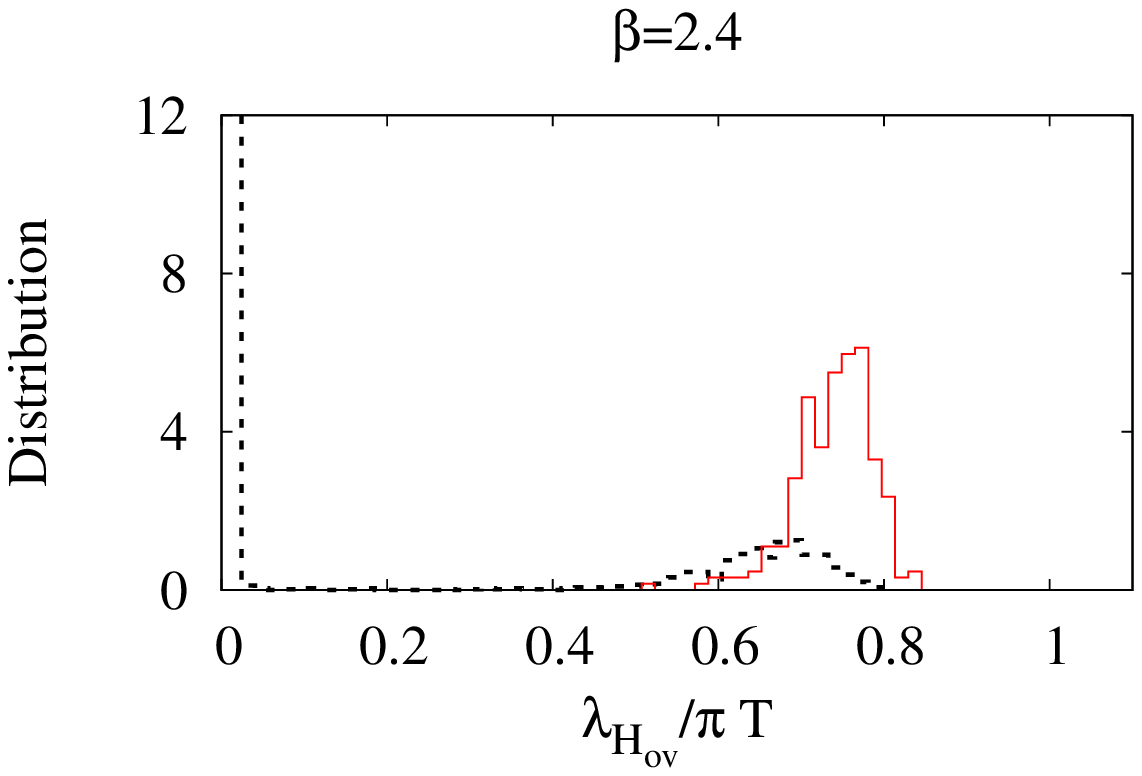}\\
  \includegraphics*[width=0.5 \textwidth,clip=true]
  {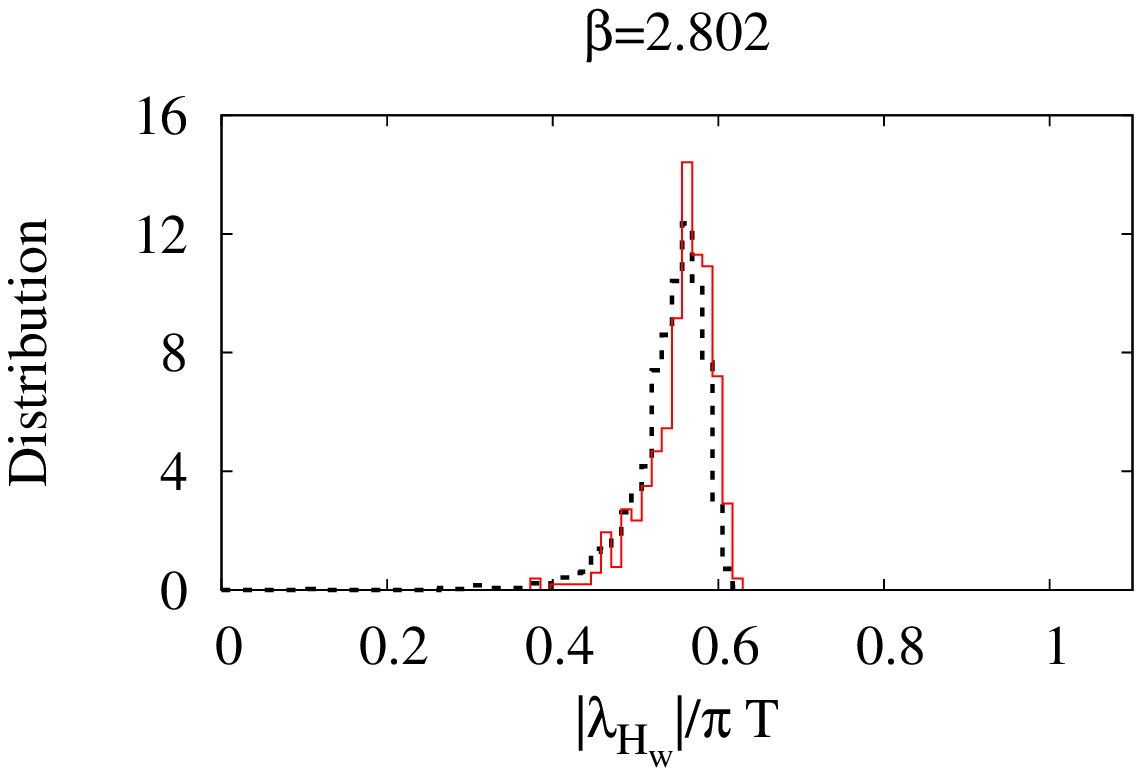}&\hspace{-4ex}
  \includegraphics*[width=0.5 \textwidth,clip=true]
  {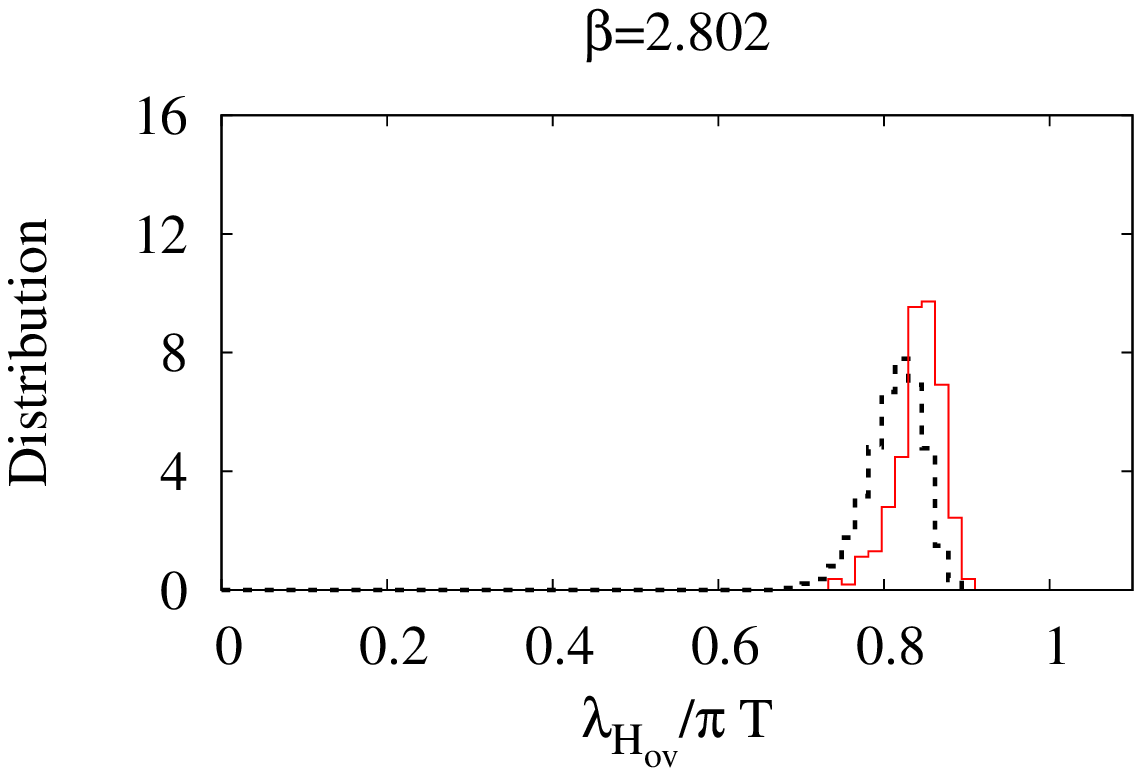}\\
  \includegraphics*[width=0.5 \textwidth,clip=true]
  {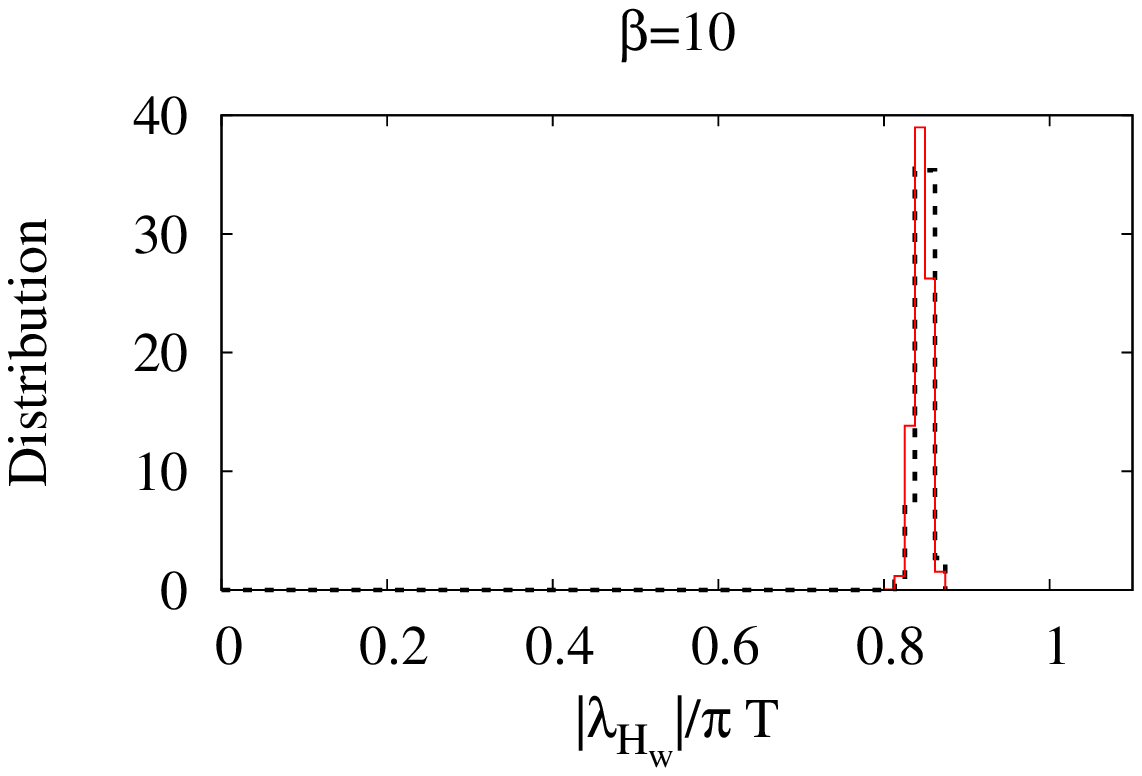}&\hspace{-4ex}
  \includegraphics*[width=0.5 \textwidth,clip=true]
  {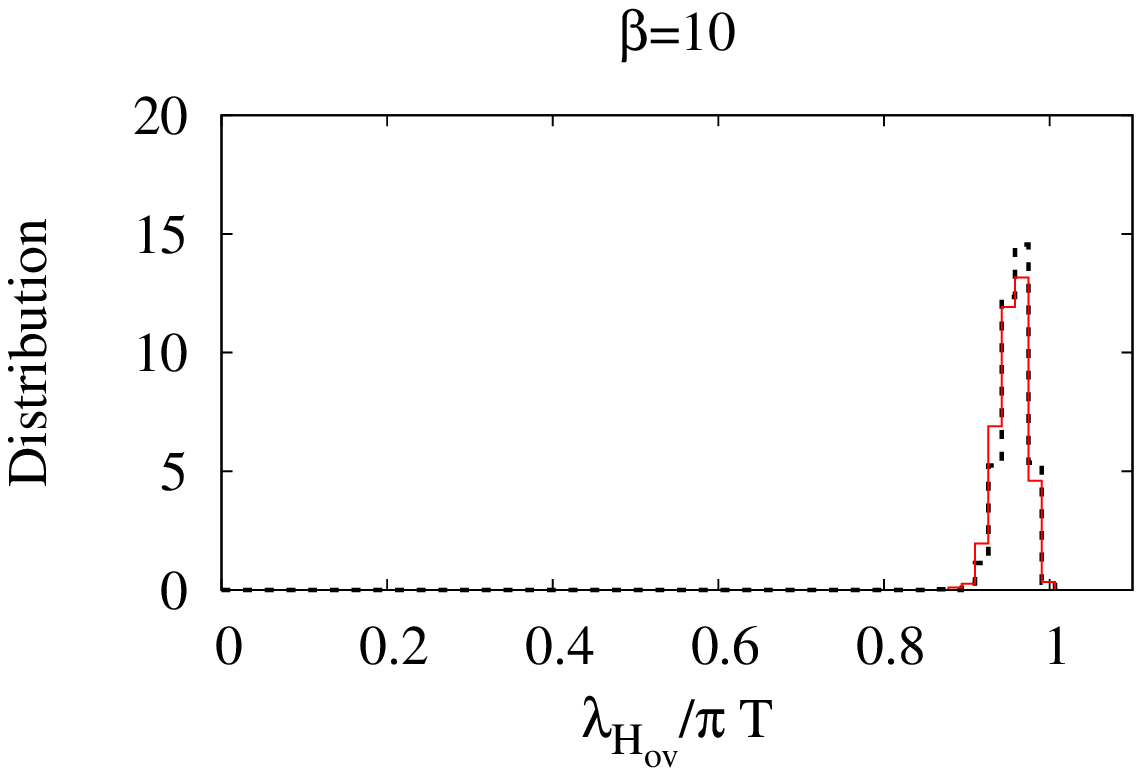}\\
  \end{tabular}
  \caption{
  Comparison of the distribution of the smallest eigenvalue on the
  configurations generated at the same $\beta$ with and without the TF
  term.
  Those of $H_{\rm W}$ (left) and $H_{\rm ov}$ (right) are shown for
  three $\beta$ values.
  }
 \label{fig:histogram_H}
 \end{center}
\end{figure}

It is also important to note that the presence of the TF term, in
general, changes the correspondence between the simulation parameter
$\beta$ and temperature $T$.
By using the fact that the spectrum of the Dirac operator is sensitive
to the temperature, we see how much the correspondence between the
simulation parameter $\beta$ and temperature $T$ is shifted in the
presence of the TF term.
The distribution of the smallest eigenvalues of the Hermitian Wilson
($H_W$) and overlap ($H_{\rm ov}$) Dirac operators are shown in
Fig.~\ref{fig:histogram_H} as examples, where 
$\beta=2.450$, $2.802$ and $10$ correspond to $T\sim 1.3\,T_c$,
$2.25\,T_c$ and $8\times 10^3\,T_c$, respectively.

In Fig.~\ref{fig:histogram_H} (left), the suppression of the appearance of
small eigenvalues is clear at $T\sim 1.3\,T_c$ (left) while no
significant difference is observed at $T > 2\,T_c$.
As for the Hermitian overlap Dirac operator
[Fig.~\ref{fig:histogram_H} (right)], while the effect of the TF term
is again clear at low temperature (top) especially in the near-zero
mode region, the distributions reasonably agree at high temperatures
(middle and bottom).
The temperature region we are interested in is $T \simge 2\,T_c$
and in such a region the Dirac spectra with and without the TF term turn
out to agree at the same $\beta$ values.
This observation allows us to employ the relationship between the
simulation parameter $\beta$ and temperature $T$ obtained in simulations
with the same gauge action but without the TF term.
Although it would be possible to numerically take the $\mu$=0 limit, we
do not pursue the limit in this exploratory study.

The configurations are generated at 12 values of $\beta$ ranging from
$T_c$ to $10^{4}\,T_c$ and in four different topological
sectors, $Q=0$, $1$, $-1$, $-2$.
The configurations are stored every 10 and 5 trajectories for
$16^3\times 4$ and $24^3\times 4$ lattices, respectively.
The simulation parameters and statistics are tabulated in
Tab.~\ref{tab:simpoints}.
The values of $T/T_c$ in the table are obtained by using the formula
provided in Ref.~\cite{Okamoto:1999hi}, where the lattice spacings are
determined in a wide range of $\beta$ using the same gauge action as
ours but without the TF term.
\begin{table}
 \begin{tabular}{rr|c|rrrr|ccc}
  $\Nsite$ & $\beta$\hspace{2ex} & $T/T_c$
  & $Q=0$ & $+1$ & $-1$ & $-2$ & $Q=+1$ & $-1$ & $-2$ 
\\
  \hline
  $16^3\times 4$
  & 2.300 &  1.02 & 1453 & 1737 & 1240 & 1207 &  $-1.3(9)$ & $-1.0(10)$ & $-2.5(10)$
\\
  & 2.400 &  1.23 & 1255 & 1352 & 1053 & 1772 & $-3.4(9)$ & $-3.2(10)$ & $-5.3(8)$
\\
  & 2.500 &  1.45 & 1490 & 1228 & 1101 & 1109 & $-1.9(8)$ & $-0.8(9)$ & $-3.1(8)$
\\
  & 2.600 &  1.69 & 1217 & 1105 & 1074 & 1229 & $-2.1(9)$ & $-1.3(9)$ & $-3.6(8)$
\\
  & 2.700 &  1.96 & 1137 & 1388 & 1344 & 1876 & $-1.5(8)$ & $-1.4(8)$ & $-3.2(8)$
\\
  & 2.802 &  2.25 & 1397 & 1338 & 1430 & 1351 & $-1.7(7)$ & $-1.8(7)$ & $-4.0(7)$
\\
  & 3.000 &  2.90 & 1876 & 1359 & 1754 & 1297 & $-1.5(6)$ & $-1.6(6)$ & $-3.5(6)$
\\
  & 3.200 &  3.70 & 1750 & 1732 & 2719 & 1204 & $-1.3(5)$ & $-0.9(5)$ & $-2.9(6)$
\\
  & 3.500 &  5.23 & 1328 & 1114 & 1100 & 1255 & $-1.4(6)$ & $-1.4(6)$ & $-3.1(5)$
\\
  & 4.000 &  9.16 & 1445 & 1197 & 1239 & 1346 & $-1.3(5)$ & $-1.3(5)$ & $-2.9(5)$
\\
  & 5.000 & 27.82 & 1097 & 1256 & 1237 & 1043 & $-1.3(4)$ & $-1.3(4)$ & $-3.0(4)$
\\
  & 10.00 & 8.2$\times 10^{3}$ & 1051 & 1054 & 1001 & 1001 & $-1.4(2)$ & $-1.6(2)$ & $-2.7(2)$
\\
  \hline
  $24^3\times 4$
  & 3.200 &  3.70 & 4152 & 3104 & 6990 & 262 & 
  $-0.4(2)$ & $-0.5(2)$ & $-1.6(7)$
\\
\end{tabular}
 \caption{Simulation parameters and the number of configurations used in
 the analysis. The rightmost three columns are $\Delta S_g^{(Q,0)}/(6\Nsite)$
 in unit of $10^{-4}$.
 }
 \label{tab:simpoints}
\end{table}

\subsection{numerical results}
\label{subsec:analysis}

\begin{figure}[htb]
 \begin{center}
  \begin{tabular}{rl}
  \includegraphics*[width=0.75 \textwidth,clip=true]
  {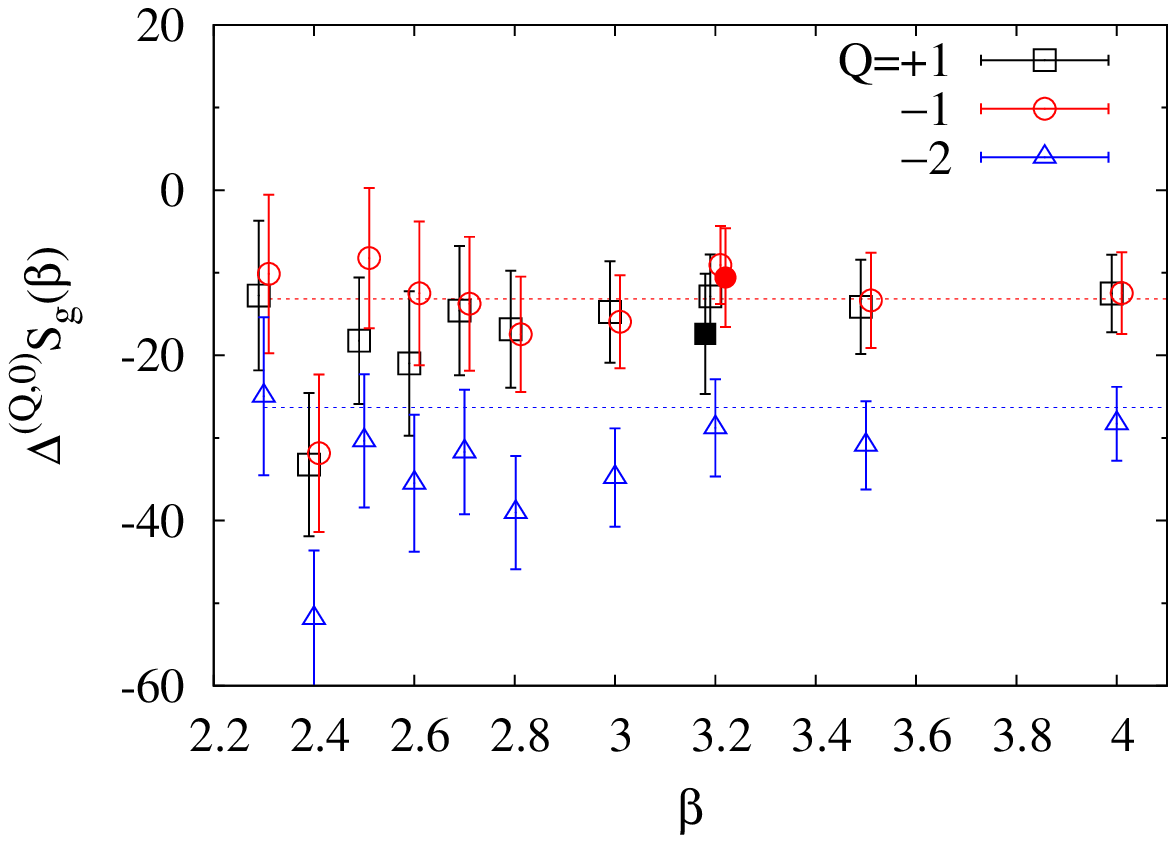} &\hspace{-16ex}
       \lower 3.4ex \hbox{
  \includegraphics*[width=0.32 \textwidth,clip=true]
  {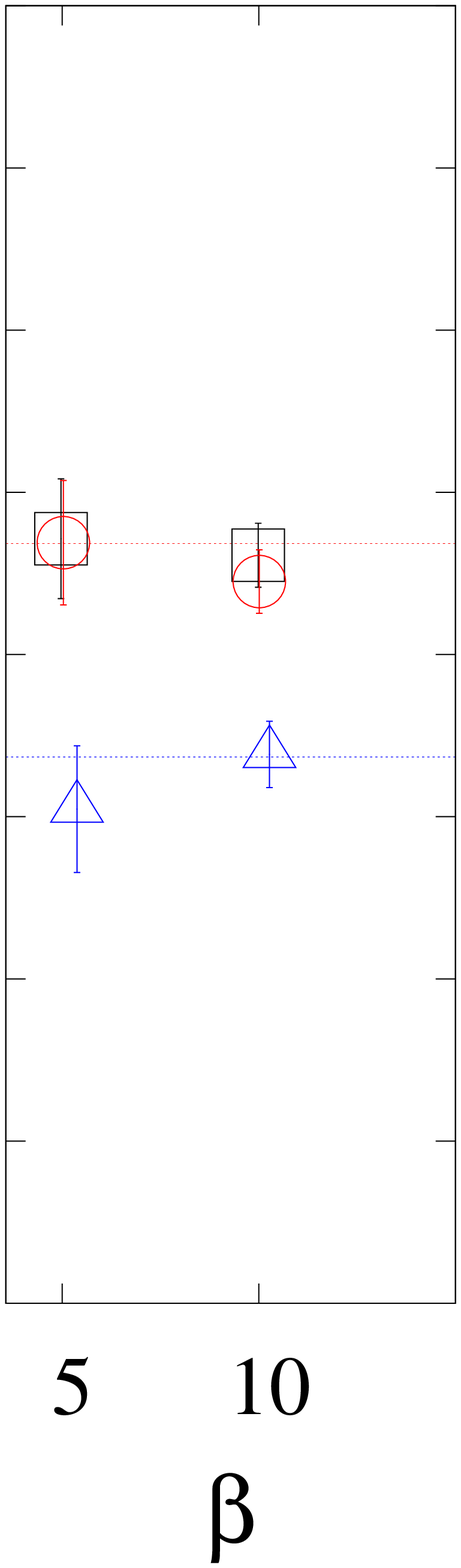}} \\[-3ex]
  \end{tabular}
 \caption{$\Delta^{(Q,0)}S_g(\beta)$ for $Q=\pm 1$ and $-2$ from
  $16^3\times 4$ (open) and $24^3\times 4$ (filled) lattices.
  The horizontal dotted lines are the high   temperature limit with
  $|Q|=1$ and $2$ from top to bottom.
  }
 \label{fig:diff-beta-dep}
\vspace{-2ex}
 \end{center}
\end{figure}

In quenched QCD, the $T$ dependence of $\chi_t$ is determined by
\begin{eqnarray}
    \frac{d\ln \chi_t(T)}{d \ln T}
&=& \frac{1}{n_{Q_2}-n_{Q_1}}
    \frac{\beta^2\,\beta_g}{6}\Delta S_g^{(Q_2,Q_1)}(\beta)
  + 4 + O(w)
\label{eq:tdep-of-chit-quench}
\ .
\end{eqnarray}
The results of $\Delta S_g^{(Q,0)}(\beta)$ with $Q=\pm 1$ and $-2$ are
shown in Fig.~\ref{fig:diff-beta-dep}, where it is seen that the data
for $Q=1$ and $-1$ agree well within the statistical error as expected.
Thus, the averaged value over $Q=1$ and $-1$ is used in the following
analysis.
\begin{figure}[htb]
\vspace*{-1ex}
 \begin{center}
  \begin{tabular}{rl}
  \includegraphics*[width=0.75 \textwidth,clip=true]
  {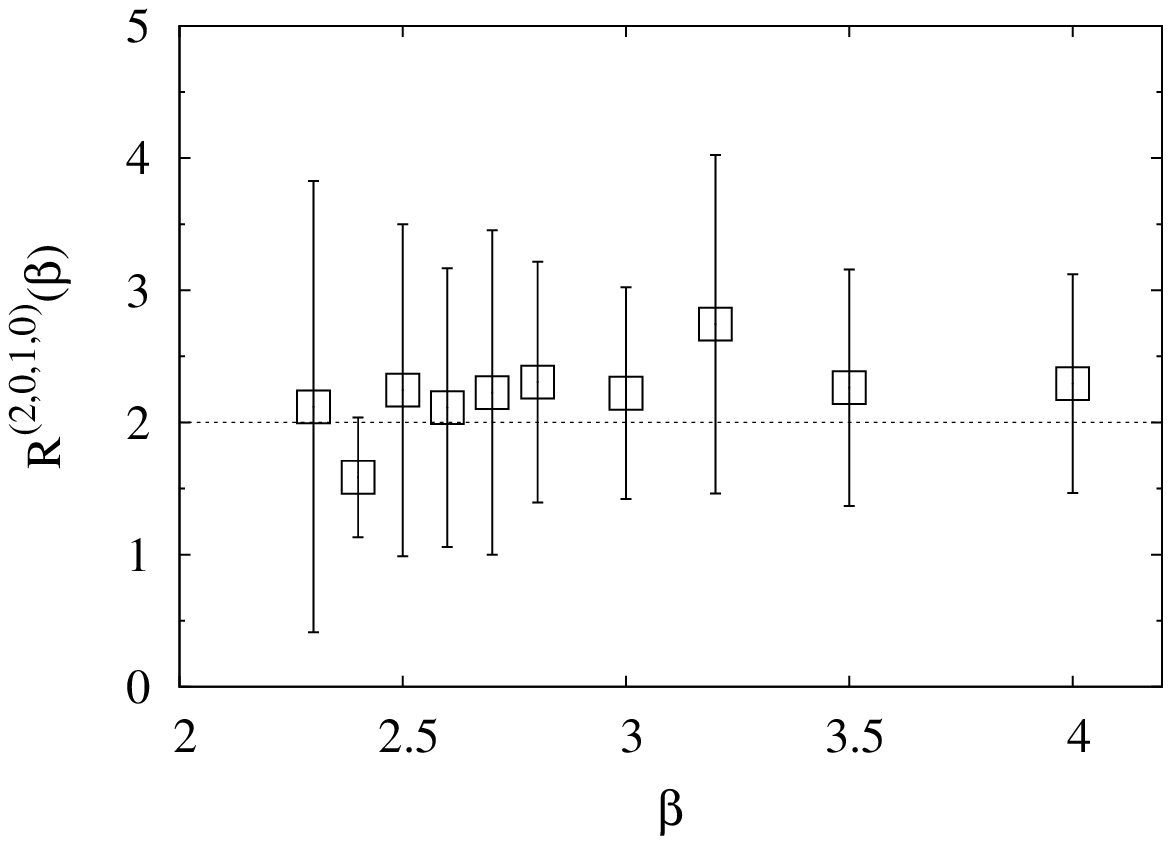}&
   \hspace{-14ex} \lower 3.4ex \hbox{
  \includegraphics*[width=0.32 \textwidth,clip=true]
  {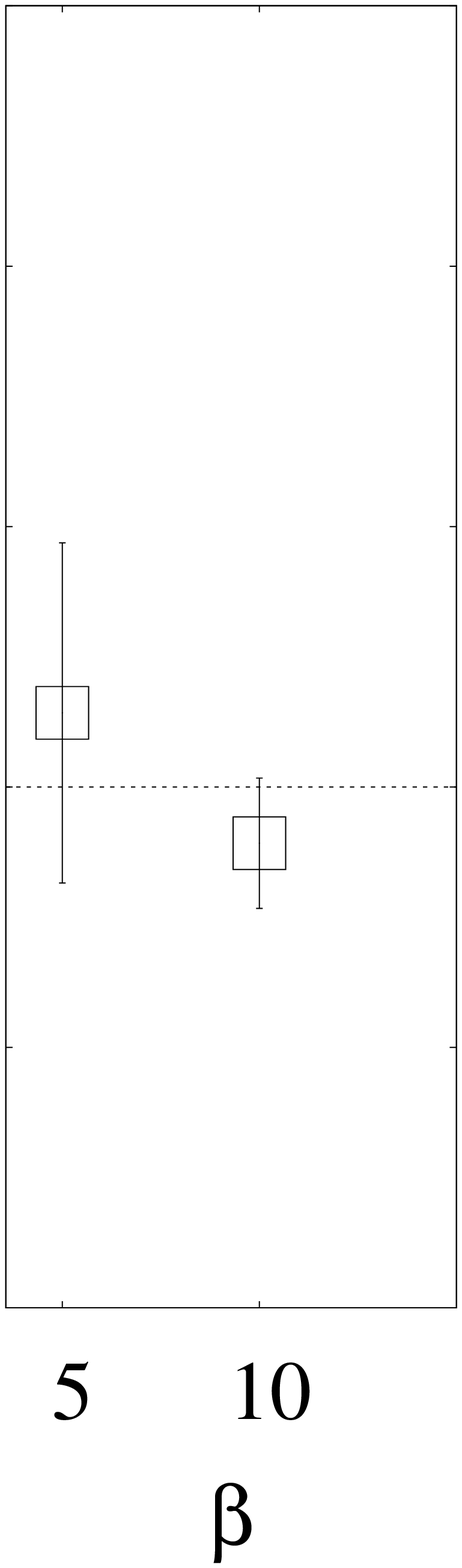}}\\[-3ex]
  \end{tabular}
 \caption{$R^{(2,0,1,0)}(\beta)$ in eq.~(\ref{eq:ratio}).
  }
 \label{fig:R}
\vspace{-4ex}
 \end{center}
\end{figure}

The horizontal dotted lines represent the action difference in the BPST
instanton solutions, or in the high temperature limit, for $|Q|=1$ and
$2$ from top to bottom.
The lattice data for $Q=\pm 1$ are on top of the corresponding BPST line
down to $\beta\sim 2.5$ (or $T\sim 1.45\,T_c$) and suddenly decrease at
$\beta\sim 2.4$.
The similar behavior is observed for $Q=-2$ but the deviation from the
corresponding BPST line starts at slightly larger $\beta$,
$\beta\sim 3$.
The jump observed at $\beta\sim 2.4$ may be associated with the phase
transition.
Studying the phase transition itself within this framework is
interesting, but we focus on the high temperature region in this paper.

The large volume results are also shown in Fig.~\ref{fig:diff-beta-dep}
(filled symbols).
It is confirmed that the $Q=\pm 1$ result are consistent with that from
the smaller lattice.
We omit the $Q=-2$ result on the larger lattice from the figure because
of a large uncertainty.

In order to estimate $d\ln\chi_t/d \ln T$ using
$\Delta^{(Q,0)}S_g(\beta)$ with $Q=\pm1$ or $-2$, we need to know
$n_{Q}-n_0$.
$n_{\pm 1}-n_0=1$ is empirically known \footnote{See the discussion
around eq.~(\ref{eq:fact}).}.
We can estimate $n_{2}-n_0$ by looking at $R^{(2,0,1,0)}(\beta)$ [see
eq.~(\ref{eq:wscaling})].
Figure~\ref{fig:R} shows that $R^{(2,0,1,0)}(\beta)$ is consistent with
two over the whole range of $\beta$ we have studied, but the large
statistical errors do not allow the precise determination except for the
region of $\beta\ge 10$.
It is seen that, when the mean value is relatively large, the error is
also large.
Thus, we assume $n_{-2}-n_0=2$ in the following analysis.

\begin{figure}[htb]
 \begin{center}
  \begin{tabular}{c}
  \includegraphics*[width=0.7 \textwidth,clip=true]
  {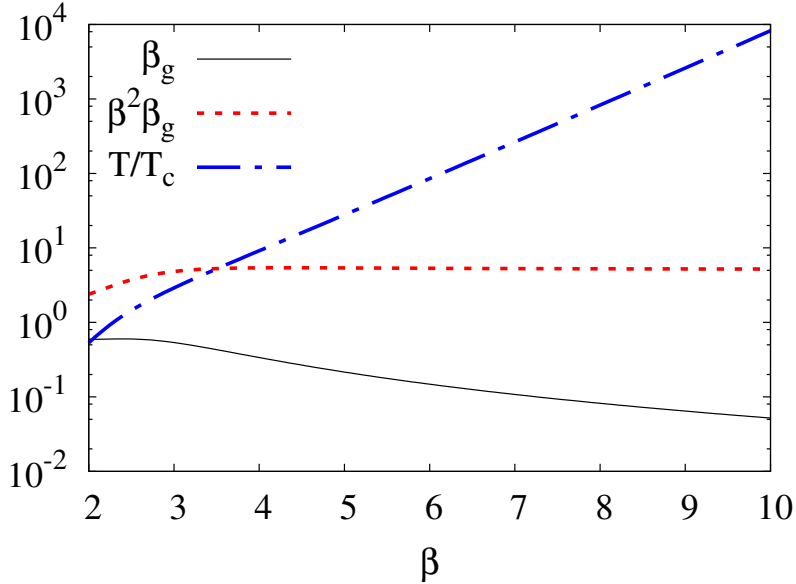}
  \end{tabular}
  \caption{$\beta$ dependence of $\beta_g$, $\beta^2\beta_g$ and
  $T/T_c$.
  }
  \label{fig:betag-beta}
 \end{center}
\end{figure}

The QCD beta function, $\beta_g$, down to a low energy scale ($\sim
T_c$) is necessary in estimating eq.~(\ref{eq:tdep-of-chit-quench}).
We use the result of Ref.~\cite{Okamoto:1999hi}, in which the lattice
spacing is expressed as a function of the lattice gauge coupling,
$\beta$, as
\begin{eqnarray}
    (a\sqrt{\sigma})(\beta)
&=& \frac{f(\beta)}{c_0}\,
    \left[ 1 + c_2 \hat a(\beta)^2 + c_4 \hat a(\beta)^4 \right]
\ ,
\end{eqnarray}
where $\sigma$ denotes the string tension and
\begin{eqnarray}
    \hat a(\beta)
= \frac{f(\beta)}{f(\beta_1)}
\ &,&\ \ \ \
    f(\beta)
= e^{-\frac{\beta}{12\,b_0}}\,
  \bigg(\frac{6\, b_0}{\beta}\bigg)^{-\frac{b_1}{2\,b_0^2}}
\ , 
\end{eqnarray}
\begin{eqnarray}
&& c_0=0.524(15),\ \ \
   c_2=0.274(76),\ \ \
   c_4=0.105(36),\ \ \
   \beta_1=2.40
\ .
\end{eqnarray}
Using this expression, $\beta_g$ is numerically determined through
\begin{eqnarray}
    \beta_g
&=& - \frac{6}{\beta^2}
      \frac{1}{\displaystyle\frac{d \ln (a\sqrt{\sigma})}{d \beta}}
\label{eq:numerical-betag}
\ .
\end{eqnarray}
At the same time, the relationship between $T/T_c$ and $\beta$ is
found to be
\begin{eqnarray}
    \frac{T(\beta)}{T_c}
&=& \frac{(a\sqrt{\sigma})(\beta_c)}{(a\sqrt{\sigma})(\beta)}
\ ,
\end{eqnarray}
where $T_c=T(\beta_c)$ and $\beta_c=2.288$~\cite{Okamoto:1999hi}.
$\beta_g$ and $T(\beta)/T_c$ are shown as a function of the lattice
gauge coupling $\beta$ in Fig.~\ref{fig:betag-beta}.
In the plot, we also show $\beta^2\beta_g$, which approaches to
$\beta^2\beta_g\to 792/(4\pi)^2\sim 5$ in the large $\beta$ limit.

\begin{figure}[tbh]
 \begin{center}
  \begin{tabular}{rl}
  \includegraphics*[width=0.75 \textwidth,clip=true]
  {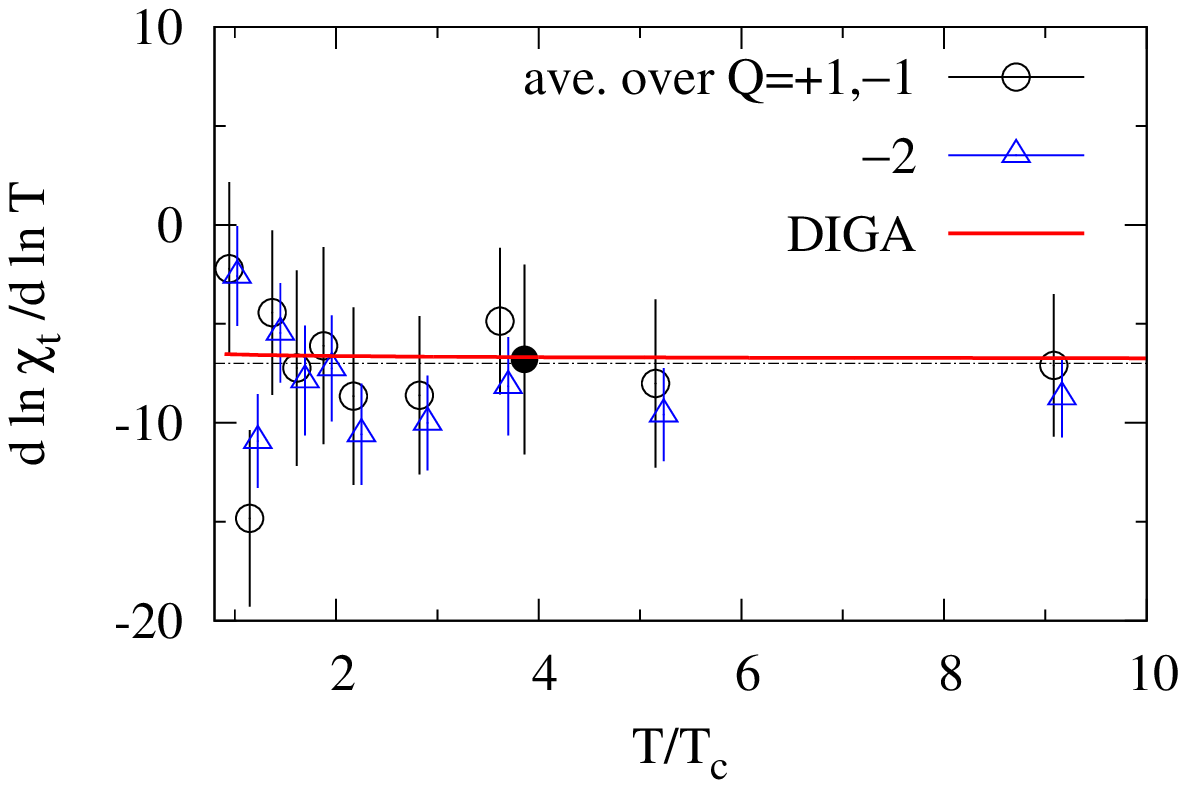} &\hspace{-14ex}
       \raise 1.4ex \hbox{
  \includegraphics*[width=0.29 \textwidth,clip=true]
  {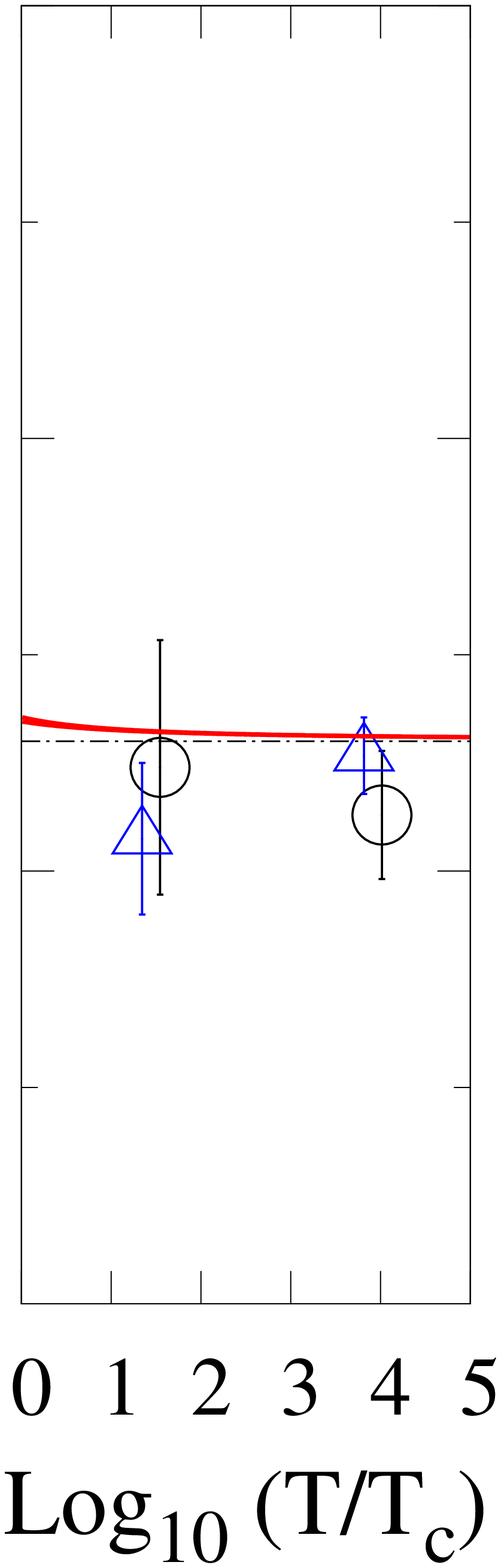}}\\
  \end{tabular}
  \caption{$T$ dependence of $\chi_t$,
  eq.~(\ref{eq:tdep-of-chit-quench}) obtained on $16^3\times 4$ (open)
  and $24^3\times 4$ (filled).
  The estimate based on the DIGA (\ref{eq:chit-diga}) with $\mu=\pi T/2$
  and $2\pi T$ (solid curves) and the high temperature limit (dashed
  line) are also shown.
  }
 \label{fig:dlnchi-dlnT-beta-dep}
 \end{center}
\end{figure}

Substituting the above results into eq.~(\ref{eq:tdep-of-chit-quench}),
$d\ln \chi_t/d\ln T$ is calculated as shown in
Fig.~\ref{fig:dlnchi-dlnT-beta-dep}, where the two solid curves represent
the prediction of the DIGA (\ref{eq:chit-diga}) with $\mu=\pi T/2$ and
$2\pi T$, respectively although they can not be distinguished at this
axis scale.

The results with $|Q|=1$ and $2$ are consistent with each other, which
is expected from the observation in Fig.~\ref{fig:R}.
These results are also consistent with the high temperature limit and
the DIGA down to $T/T_c\sim 1.5$.
Note that the results using the $Q=-2$ sector has the uncertainty smaller
than those using $Q=\pm 1$ by a factor $n_{-2}-n_0=2$, which indicates
that once the $n_{Q}-n_0$ has been fixed one can obtain very accurate
result by performing a simulation at large $Q$.

One of the concerns in this approach is the finite volume effect since
the physical volume becomes extremely small at large $\beta$.
Figure~\ref{fig:dlnchi-dlnT-beta-dep} shows that the lattice results
well reproduce the high temperature limit at high temperature.
From this observation, it is unlikely that the finite size effect
significantly affects the lattice results, and it is natural to think
that $N_T \ll N_S$ is the necessary condition for the finite volume
effects to be under control.
Indeed, the aspect ratio of our lattices is $N_S/N_T=4$, and hence the
above condition seems to be satisfied.
Nevertheless, calculations with different lattice sizes are clearly
useful to explicitly check the finite size effect and whether $w\ll 1$
holds or not.
However, since the uncertainty of the action value grows as
$\sqrt{\Nsite}$, we need the statistics proportional to $\Nsite$ to keep
the size of the uncertainty constant.

From the phenomenological point of view, $\chi_t(T)$ for 
$T_c \simle T \simle 10\,T_c$ is important.
In this range of $T$, the statistical uncertainty is relatively large
(typically $\pm 4$ for $O(10,000)$ trajectories), which makes the axion
abundance ambiguous.
It is thus important to accumulate a large number of statistics.
On the other hand, if $\chi_t(T)$ behaves like a step function, our
method should be able to detect such a behavior.

\section{summary and future prospects}
\label{sec:summary}

The QCD topological susceptibility, $\chi_t$, at high temperature
provides an important input for the estimate of the axion abundance in
the present universe.
Existing methods to calculate $\chi_t$ on the lattice in the
literature fail when $\chi_t(T)V_4\ll 1$.
We proposed a novel lattice method to calculate the temperature
dependence of the susceptibility, which is expected to work well
especially in high temperature region where $\chi_t(T)V_4\ll 1$.
To see how it works, we performed quenched simulations on the
$16^3\times 4$ lattice, and found that the results of
$d\ln \chi_t/d\ln T$ well agree with the DIGA prediction above
$1.5\, T_c$.
The simulation on a slightly larger lattice confirms that there is no
unexpected large finite volume effect, although keeping the statistical
error constant requires statistics proportional to $\Nsite$.
Thus, that error may be the main source of uncertainty in future serious
works.

To predict the axion abundance, we still have to include dynamical
quarks with the physical masses.
In order for the method to work, the difference of the chiral condensate
between two topological sectors has to be precisely determined, for
which the dynamical overlap fermion seems to be preferred.
Then, accumulating a large number of configurations requires large
amount of resources.
But, if $\chi_t(T)$ behaves like a step function, a large number of
statistics may not be necessary to detect such a behavior.

A possible way out is to generate configurations in large $Q$ sectors,
with which one can achieve an uncertainty smaller than that with $|Q|=1$
by a factor of $n_{Q}-n_0$.
Note that this requires the signal on $R^{(Q,0,1,0)}$
[eq.~(\ref{eq:ratio-3})] at $w \ll 1$ to be sufficiently precise to
unambiguously identify the integer $n_Q-n_0$.
Knowing $n_Q-n_0$ for various $Q$ is also useful to put the
constraints on the $\theta$ dependence of $E(\theta)$ in
eq.~(\ref{eq:ztheta}), whose general form would be given by
\begin{eqnarray}
  E(\theta)=\sum_n c_n(1-\cos (n\, \theta))
\label{eq:fn4}
\ ,
\end{eqnarray}
with $\chi_t=\sum_n c_n/n^2$.

\section*{Acknowledgement}

This work is in part based on Bridge++ code
(http://bridge.kek.jp/Lattice-code/).
This work is supported by JSPS KAKENHI Grant-in-Aid for Scientific
Research (B) (No.~15H03669 [JF, RK, NY]) and MEXT KAKENHI Grant-in-Aid
for Scientific Research on Innovative Areas (No.~25105011 [RK]), and
also by the Large Scale Simulation Program No.~15/16-21 of High Energy
Accelerator Research Organization (KEK).

\appendix

\section{quark actions}
\label{sec:dynamical}

The Wilson and overlap actions mentioned in the main text are described
below.
The Wilson quark action is given by
\begin{eqnarray}
    S^{\rm W}_q(\bar m_q)
&=& \sum_{f=1}^{N_f}\sum_{x,y}
     \bar {q_f}_x\, D_W(\bar m_q)_{x,y}\, {q_f}_y
\ ,
\end{eqnarray}
where
\begin{eqnarray}
    {D_W}(\bmq)_{x,y}
= \big(\bmq + 4 \big) \delta_{x,y}
  - \frac{1}{2}\sum_\mu
     \bigg\{(1-\gamma_\mu)U_\mu(x         )\delta_{x+\hat\mu,y}
          +(1+\gamma_\mu)U^\dag(x-\hat\mu)\delta_{x-\hat\mu,y}
     \bigg\}
\ ,
\label{eq:Dw}
\end{eqnarray}
where $U_\mu(x)$ is the link variable in the $\mu$ direction.
The Hermitian Wilson Dirac operator appearing in eq.~(\ref{eq:topfix})
is then given by
\begin{eqnarray}
    H_W(\bmq)
&=& \gamma_5 D_W(\bmq)
\ .
\label{eq:Hw}
\end{eqnarray}
Note that, when using the Hermitian Wilson Dirac operator in the
topology fixing term and the kernel of the overlap Dirac operator (see
below), the mass has to be negative.

We adopt the fermionic definition based on the index theorem to estimate
the topological charge for given configurations, which requires the
number of zero modes of the overlap Dirac
operator~\cite{Neuberger:1998wv}.
The overlap quark action and Dirac operator is given by
\begin{eqnarray}
    S^{\rm ov}_q(\bar m_q)
&=& \sum_{f=1}^{N_f}\sum_{x,y}
     \bar {q_f}_x\, D^{\rm ov}(\bar m_q)_{x,y}\, {q_f}_y
\ ,\\
    D^{\rm ov}(\bar m_q)_{x,y}
&=&  D^{\rm ov}(0)_{x,y}
   + \bar m_q \bigg(\delta_{x,y}
   - \frac{1}{2M_0}D^{\rm ov}(0)_{x,y}\bigg)
\label{eq:Dov}
,\\
    D^{\rm ov}(0)_{x,y}
&=& M_0\,\bigg[\, 1+ \gamma_5\, {\rm sign}(H_W(-M_0))\,\bigg]_{x,y}
\ .
\label{eq:massless-Dov}
\end{eqnarray}
We calculated the low-lying eigenvalues of the Hermitian overlap Dirac
operator, $\gamma_5D^{\rm ov}(0)$, and count the number of left- and
right-handed zero modes.
With the above definitions, it is straightforward to derive
eqs.~(\ref{eq:qqbar-def-wil}) and (\ref{eq:qqbar-def}).

\section{the $w\gg 1$ case}
\label{sec:wgg1}

The relationship between expectation values in the $\theta$ vacuum and
a fixed topological sector is derived in
Refs.~\cite{Brower:2003yx,Aoki:2007ka,Dromard:2014ela}, where the
results are written in the form of $1/w$ expansion ($w=\chi_tV_4$).
From eq.~(2.38) of Ref.~\cite{Dromard:2014ela}, it is read that
\begin{eqnarray}
    \frac{d\ln\frac{Z_{Q_2}}{Z_{Q_1}}}{d\ln w}
&=& \frac{Q_2^2-Q_1^2}{2w}
    \Bigg(
           1
	 + \frac{1}{2\chi_t}\frac{d c_4}{dw}
         - \frac{3\,(c_4/\chi_t)}{2w}
         - \frac{1}{12\chi_t\,w}\frac{d c_4}{dw}
    \Bigg)
  + O(1/w^3)
\ ,
\no\\
\label{eq:wgg1}
\end{eqnarray}
where $c_{2n}$ is defined as the coefficient of $\theta^{2n}$ in the
expansion,
\begin{eqnarray}
    E(\theta)
&=& \sum_{n=1}^\infty \frac{c_{2n}(T)}{(2n)!}\theta^{2n}
\ ,
\end{eqnarray}
especially $c_2(T)=\chi_t(T)$.
Note that $c_i(T)$ is dynamical quantity and depends on $T$.
No assumption is made in deriving eq.~(\ref{eq:wgg1}) except for
$w\gg 1$ and $Q^2 \ll w$, but in order for the expansion to be sensible,
$c_{2n}/\chi_t\sim O(1)$ is required.
It is interesting to see that eq.~(\ref{eq:wgg1}) is proportional to
$Q_2^2-Q_1^2$ while the same quantity is to $|Q_2|-|Q_1|$ when $w\ll 1$
[see, for example, eq.~(\ref{eq:dzdw-small-w})].

Using eq.~(\ref{eq:wgg1}) and estimating
$d\ln(Z_{Q_2}/Z_{Q_1})/d\ln T$, one can estimate the $T$ dependence
of $\chi_t$ through eq.~(\ref{eq:dzdt_vs_dwdt}).
Although $d\ln(Z_{Q_2}/Z_{Q_1})/d\ln T$ can be directly estimated on
the lattice as explained in the main text, we can further use the
relationship between expectation values in the $\theta$ vacuum and a
fixed topological
sector~\cite{Brower:2003yx,Aoki:2007ka,Dromard:2014ela} to proceed.
From eq.~(3.20) in Ref.~\cite{Dromard:2014ela}, it follows that
\begin{eqnarray}
    \langle O \rangle^{(Q_2)}-\langle O \rangle^{(Q_1)}
&=& O_0
    \frac{1}{2w^2}
    \Bigg[ x_2 + \frac{x_4-3(c_4/\chi_t)x_2-3\,x_2^2}{2w}
    \Bigg]\,(Q_2^2-Q_1^2)
  + O(1/w^4)
\label{eq:opwgg1}
\ ,
\end{eqnarray}
where $x_k$ are defined as follows,
\begin{eqnarray}
      x_{2k}
&=&  - \frac{d^{2k} \ln\left(\sum_{k=0}^{\infty}
	                     \frac{O_{2k}}{O_0\,(2k)!}\theta^{2k}
		       \right)}
            {d\theta^{2k}}\bigg|_{\theta=0}
\\
      x_0
&=&   0
\\
      x_2
&=&  - \frac{O_2}{O_0}
 =  - w\bigg( 1 - \frac{\langle Q^2 O\rangle_{\theta=0}}{O_0\,w}\bigg)
\\
      x_4
&=&  - \frac{O_4}{O_0}
    + 3\left(\frac{O_2}{O_0}\right)^2
 =  - 6\,w^2\left(1- \frac{\langle Q^2 O \rangle_{\theta=0}}{O_0\,w}\right)
    + \left(q_4 - \frac{\langle Q^4 O \rangle_{\theta=0}}{O_0}\right)
    + 3\, x_2^2
\ .
\end{eqnarray}
$O_{2n}$ is defined by 
\begin{eqnarray}
      \langle O \rangle_{\theta}
&=&   \sum_{n=0}^{\infty} \frac{O_{2n}}{(2n)!}\theta^{2n}
\ ,
\end{eqnarray}
and $q_4=\langle Q^4 \rangle_{\theta=0}$.
The $1/w$ expansion in eq.~(\ref{eq:opwgg1}) is sensible if $x_i\sim O(1)$.

Applying the above expansions to the quenched case yields
\begin{eqnarray}
    \frac{d\ln \chi_t(T)}{d \ln T}
&=& - \frac{\beta\beta_g}{6}
      \langle S_g \rangle_{\theta=0}
      \frac{x_2}{pw}
      \Bigg[ 1 + \frac{(x_4/x_2) - (c_4/\chi_t) - 3x_2}{2w}
               + \frac{3\,c_4/(\chi_t p)}{2w}
               + \frac{1/(\chi_t p)}{12w}\frac{d\,c_4}{dw}
      \Bigg]
\no\\&&\hspace{0ex}
       + 4 + O(1/w^3)
\ ,
\label{eq:tdep-of-chit-large}
\end{eqnarray}
where
\begin{eqnarray}
    p
&=& 1 + \frac{1}{2\chi_t}\frac{d\,c_4}{dw}
\ .
\end{eqnarray}

\section{hopping parameter expansion}
\label{sec:hpe}

We consider the temperature dependence of $\chi_t$ in the presence of
$N_f$ flavors of heavy quarks (and no light quarks).
In the following, we assume that the degenerate heavy mass is larger
than the temperature, $m_q\gg T$.
By introducing the hopping parameter
\begin{eqnarray}
    \kappa_q
&=& \frac{1}{2(\bmq+4)}
\label{eq:kappa-def}
\ ,
\end{eqnarray}
the Wilson Dirac operator, eq.~(\ref{eq:Dw}), can be rewritten as
\begin{eqnarray}
    {D_W}(\bmq)_{x,y}
&=& \big(\bmq + 4 \big) \delta_{x,y}
  - \frac{1}{2}\sum_\mu
     \bigg\{(1-\gamma_\mu)U_\mu(x         )\delta_{x+\hat\mu,y}
          +(1+\gamma_\mu)U^\dag(x-\hat\mu)\delta_{x-\hat\mu,y}
     \bigg\}
\no\\
&=& \frac{1}{2\kq}
    \Bigg[\delta_{x,y}
  - \kq\sum_\mu
     \bigg\{(1-\gamma_\mu)U_\mu(x         )\delta_{x+\hat\mu,y}
          +(1+\gamma_\mu)U^\dag(x-\hat\mu)\delta_{x-\hat\mu,y}
     \bigg\}\Bigg]
\ .
\no\\
\end{eqnarray}
Thus, when $\kq\ll 1$, we can expand $D_W$ in terms of $\kq$.

If we take the Wilson fermion as the heavy quark action, the partition
function and the expectation value of operator $O$ in a fixed topology
sector can be written as
\begin{eqnarray}
    Z_Q(\beta,\bmq)
&=&
    \int_{\in Q} \!\! {\cal D} U\,e^{- S_g(\beta)}\, 
    \big[\det D_W(\bmq)\big]^{N_f}
\ ,
\\
    \langle \rangle_{\beta,\bmq}^{(Q)}
&=&
    \frac{1}{Z_Q(\beta,\bmq)}
    \int_{\in Q} \!\! {\cal D} U\,e^{- S_g(\beta)}\, 
    \big[\det D_W(\bmq)\big]^{N_f}
    O
\label{eq:expctv-w}
\ ,
\end{eqnarray}
respectively.
Applying the hopping parameter expansion (HPE) to the heavy quarks, the
determinant can be expanded as
\begin{eqnarray}
   \left[ \det D_W (\bmq) \right]^{\Nf}
&=& \bigg(\frac{1}{2\kq}\bigg)^{12\,N_f\,\Nsite}
    \exp\left[ \Nf\,\kq^4\,Y \right] + O(\kq^6)
\label{eq:detmw-1}
\ ,
\\
    Y
&=& 48\,\Nsite\,(6\, W_P + L)
\label{eq:Y}
\ .
\end{eqnarray}
where $L$ denotes the real part of the Polyakov loop and we have used
$N_T=4$.
Then, the expectation value, eq.~(\ref{eq:expctv-w}), is given by
\begin{eqnarray}
    \langle O \rangle_{\beta,\bmq}^{(Q)}
&=&   \frac{\langle e^{N_f\,\kq^4\, Y}  O \rangle^{(Q)}_{\beta}}
           {\langle e^{N_f\,\kq^4\, Y} \rangle^{(Q)}_{\beta}}
    + O(\kq^6)
\ .
\end{eqnarray}

If $O$ consists of quark fields like $O=\bar q_xq_x$, the HPE is further
applied.
Using
\begin{eqnarray}
      {\rm Tr} \big[ D_W^{-1} \big]
&=&    8\kq\,\left( 3\, \Nsite - \kq^4\,Y \right)
    + O(\kq^6)
\ ,
\end{eqnarray}
$\bmq\langle s_{\bar qq} \rangle^{(Q)}_{\beta,\bmq}$ can be
written as
\begin{eqnarray}
    \bmq\langle s_{\bar qq} \rangle^{(Q)}_{\beta,\bmq}
&=& \bmq \sum_x
    \langle \bar q_x q_x \rangle^{(Q)}_{\beta,\bmq}
 =  - \bmq \sum_x
      \frac{\langle\, (D_W)^{-1}_{x,x}\ [\,\det D_W(\bmq)\,]^{N_f}\,
            \rangle^{(Q)}_{\beta}}
           {\langle\, [\,\det D_W(\bmq)\,]^{N_f}\,
            \rangle^{(Q)}_{\beta}}
\no\\
&=& - 12\,\left(1 -8\kq\right)\bigg[\Nsite - \frac{\kq^4}{3}
      \frac{\langle\, Y\,
                      e^{N_f\,\kq^4\, Y}
            \rangle^{(Q)}_{\beta}}
	    {\langle\, e^{N_f\,\kq^4\, Y} \rangle^{(Q)}_{\beta}}
	    \bigg] + O(\kq^6)
\ .
\end{eqnarray}
In summary,
\begin{eqnarray}
    \frac{d\ln \chi_t(T)}{d \ln T}
&=& \frac{1}{n_{Q_2}-n_{Q_2}}
    \Bigg[
    - \frac{\beta\,\beta_g}{6}
      \bigg(
          \frac{\langle S_g\, e^{N_f\kq^4 Y}\rangle_{\beta}^{(1)}}
               {\langle e^{N_f\kq^4 Y}\rangle_{\beta}^{(1)}}
        - \frac{\langle S_g\, e^{N_f\kq^4 Y}\rangle_{\beta}^{(0)}}
            {\langle  e^{N_f\kq^4 Y} \rangle_{\beta}^{(0)}}
      \bigg)
\no\\&&
    +\, 4\,N_f\,\kq^4\,\left(1 -8\kq\right)\,
      \bigg( 1 + \frac{d \ln m_q}{d\ln a} \bigg)\,
      \bigg(
         \frac{\langle\, Y\,e^{N_f\,\kq^4\, Y}
	       \rangle^{(1)}_{\beta}}
              {\langle\, e^{N_f\,\kq^4\, Y}
	       \rangle^{(1)}_{\beta}}
       - \frac{\langle\, Y\,e^{N_f\,\kq^4\, Y}
               \rangle^{(0)}_{\beta}}
	      {\langle\, e^{N_f\,\kq^4\, Y}
	      \rangle^{(0)}_{\beta}}
     \bigg)\Bigg]
\no\\&&
    +\, 4 + O(\kq^6)
\label{eq:tdep-of-chit-hpe-2}
\ .
\end{eqnarray}
It turns out that, in the heavy quark region, the Polyakov loop plays an
important role.

\begin{figure}[h]
 \begin{center}
  \begin{tabular}{rl}
  \includegraphics*[width=0.70 \textwidth,clip=true]
  {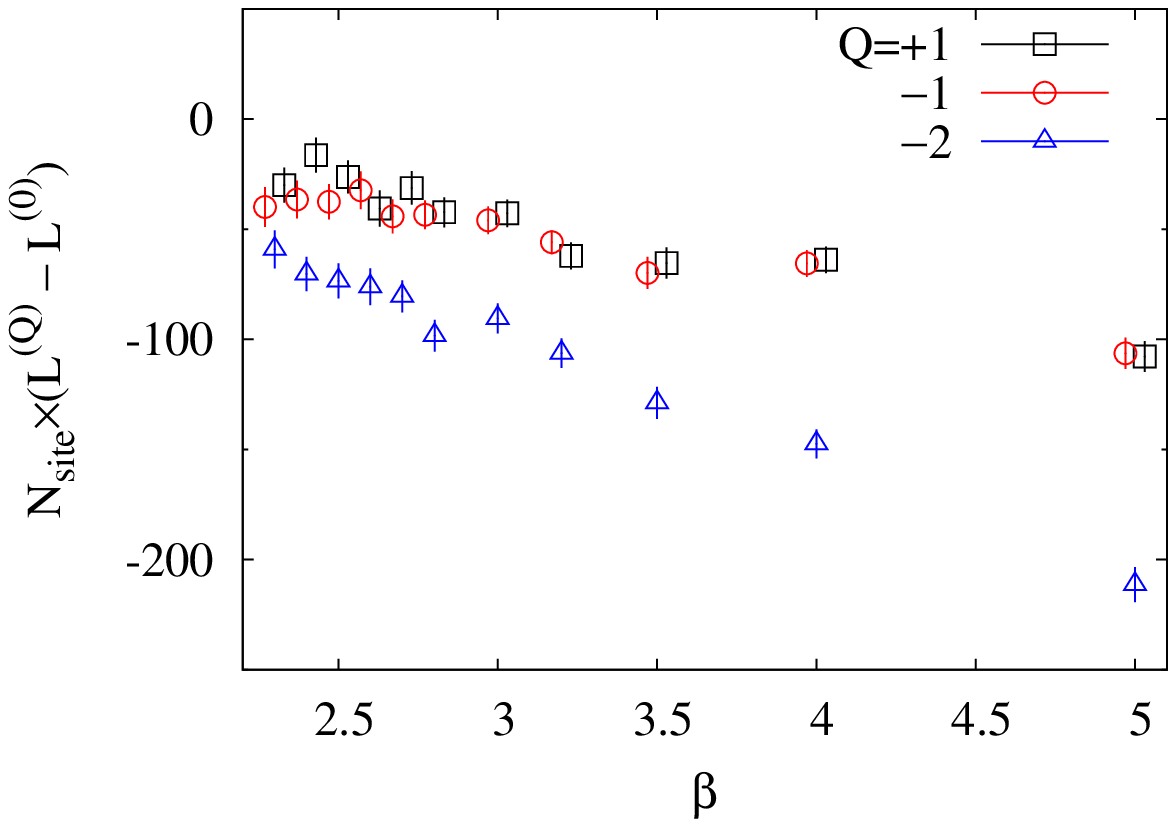} &\hspace{-8ex}
  \lower 3.3ex \hbox{
  \includegraphics*[width=0.3 \textwidth,clip=true]
  {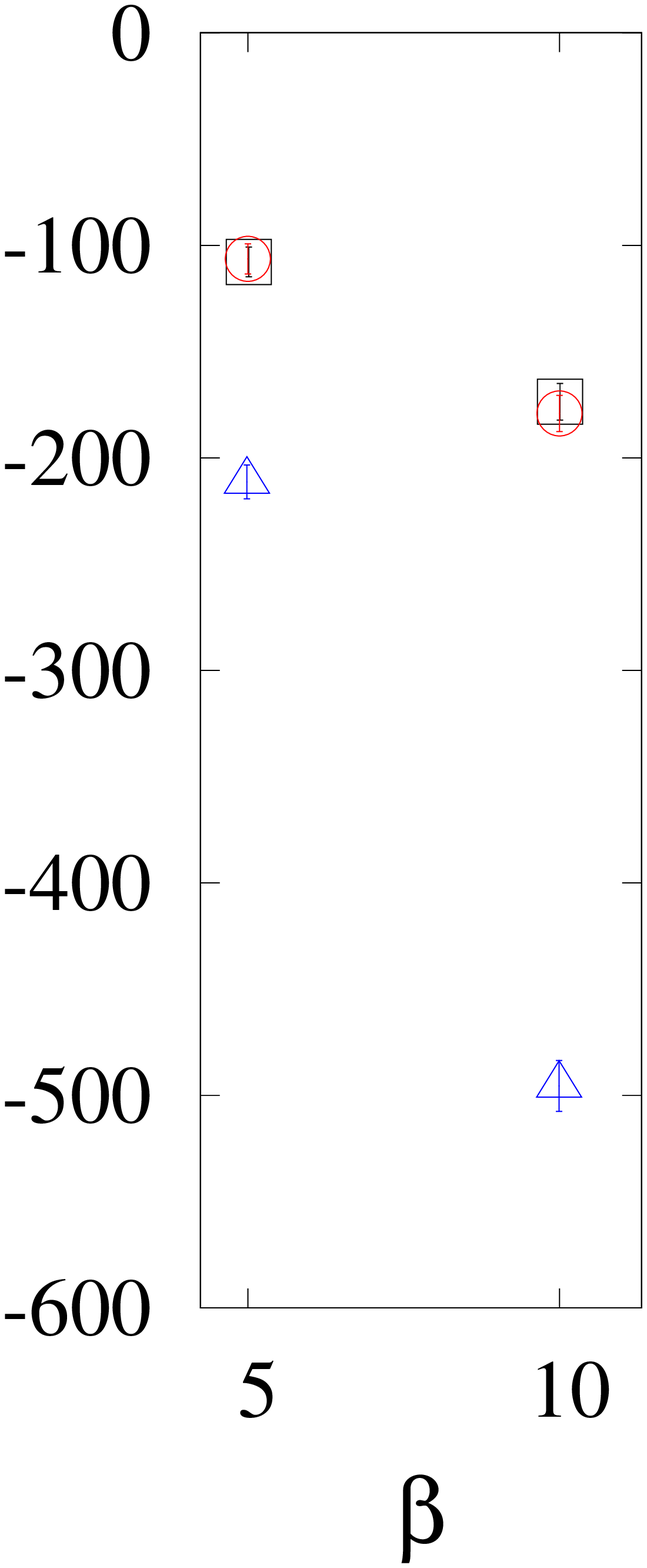}} \\
  \end{tabular}
\vspace{-2ex}
 \caption{$\beta$ dependence of
  $\Nsite\times(L^{(Q)}-L^{(0)})$.}
 \label{fig:ploop-diff-beta-dep}
 \end{center}
\end{figure}

In Fig.~\ref{fig:ploop-diff-beta-dep}, the difference of $L$ in
different $Q$ sectors is shown.

\begin{figure}[h]
 \begin{center}
  \begin{tabular}{cc}
  \hspace{-5ex}
  \includegraphics*[width=0.52 \textwidth,clip=true]
  {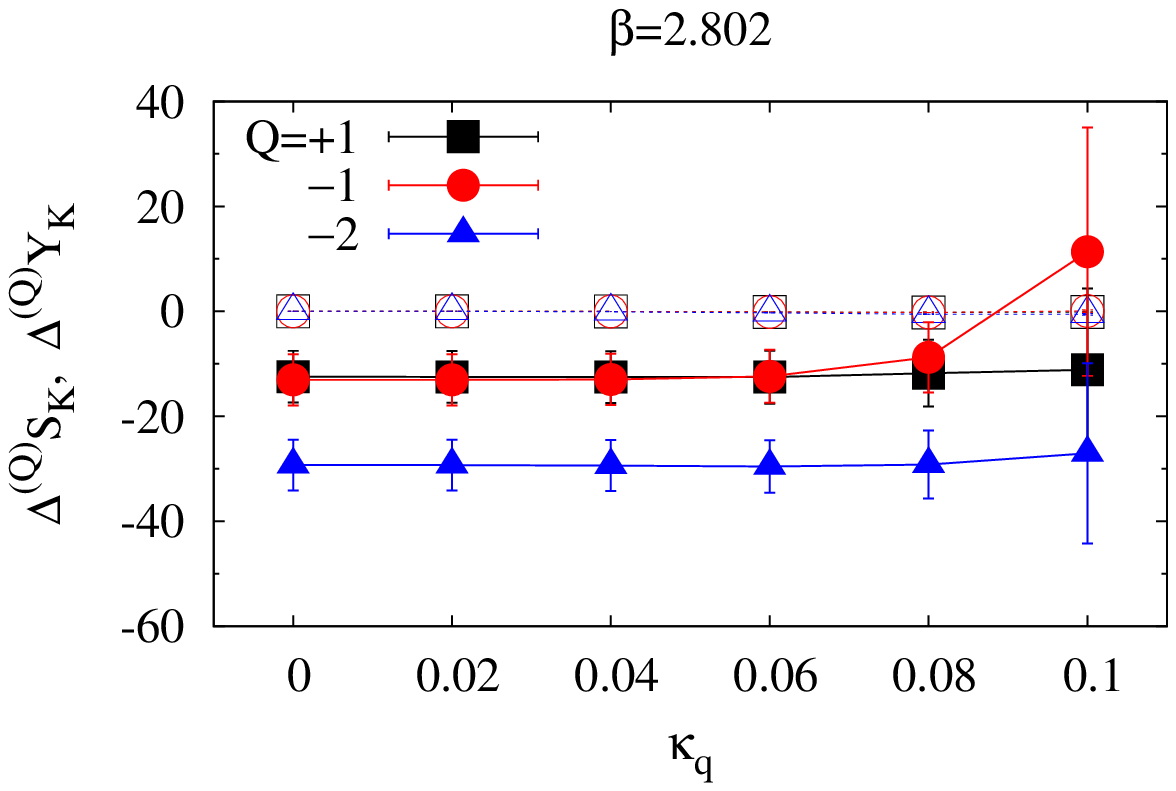} &
  \hspace{-4ex}
  \includegraphics*[width=0.52 \textwidth,clip=true]
  {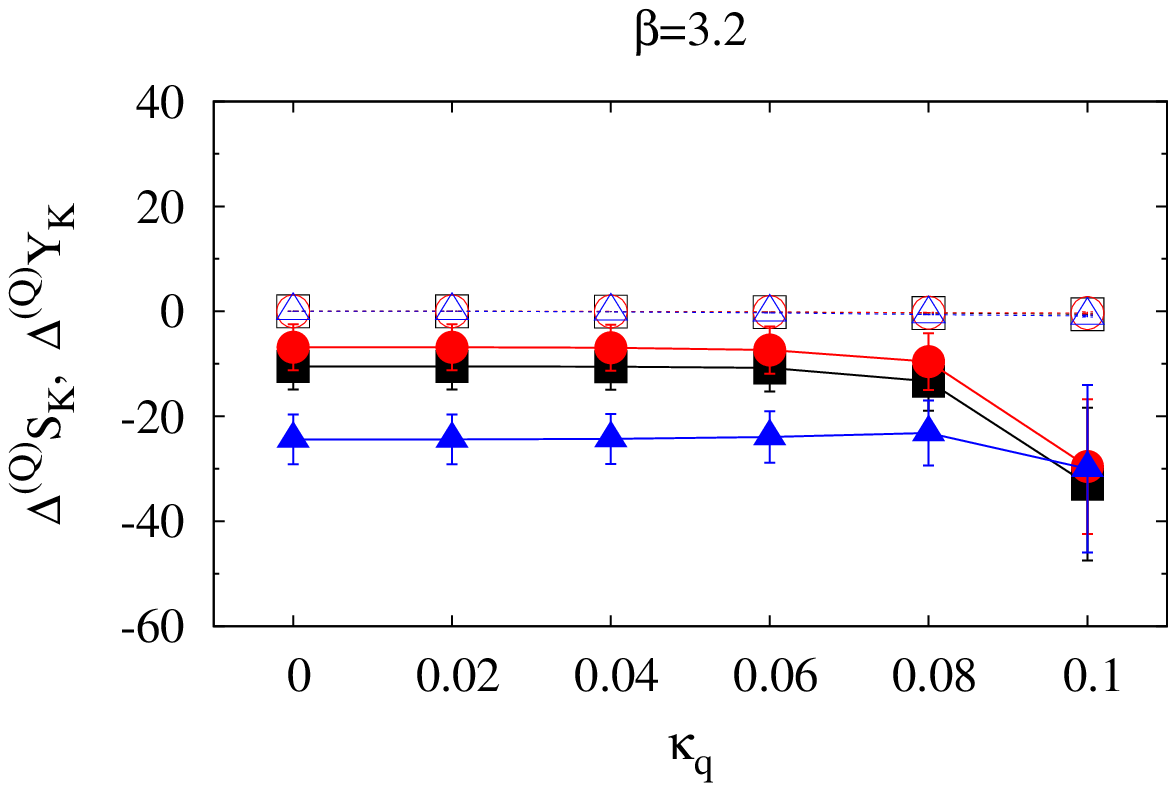} \\
  \hspace{-5ex}
  \includegraphics*[width=0.52 \textwidth,clip=true]
  {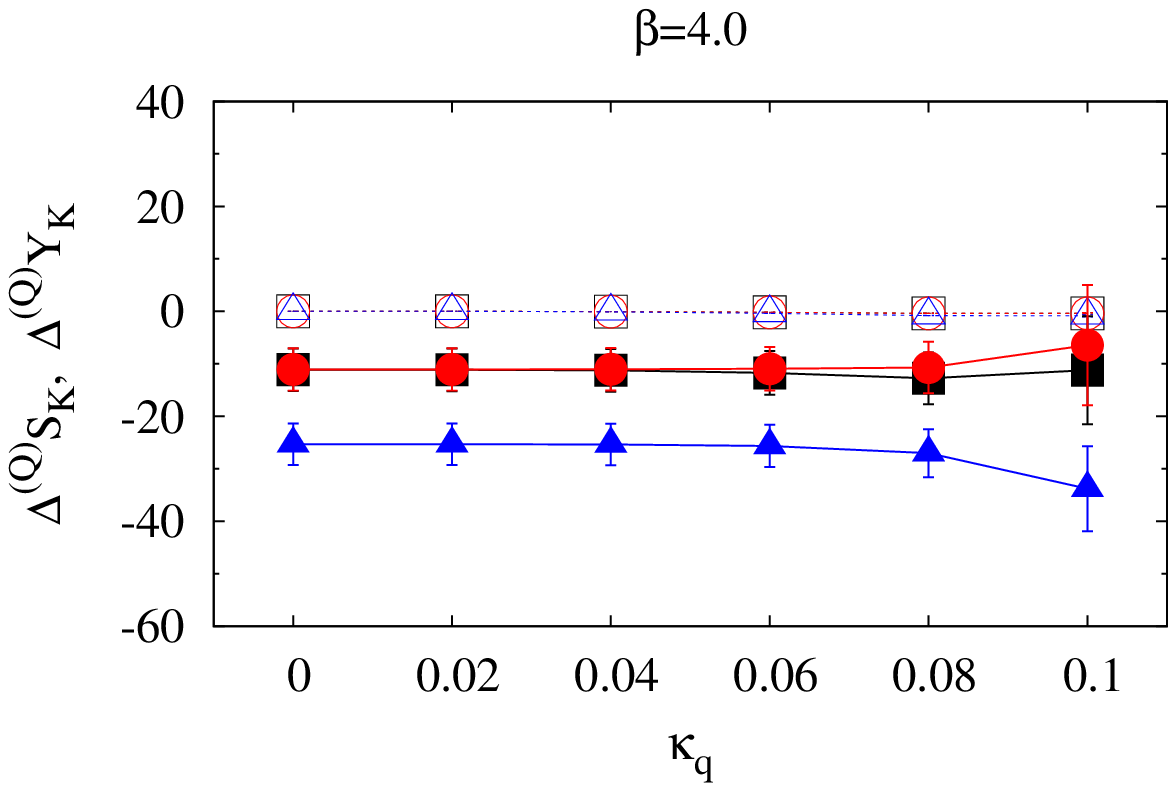} &
  \hspace{-4ex}
  \includegraphics*[width=0.52 \textwidth,clip=true]
  {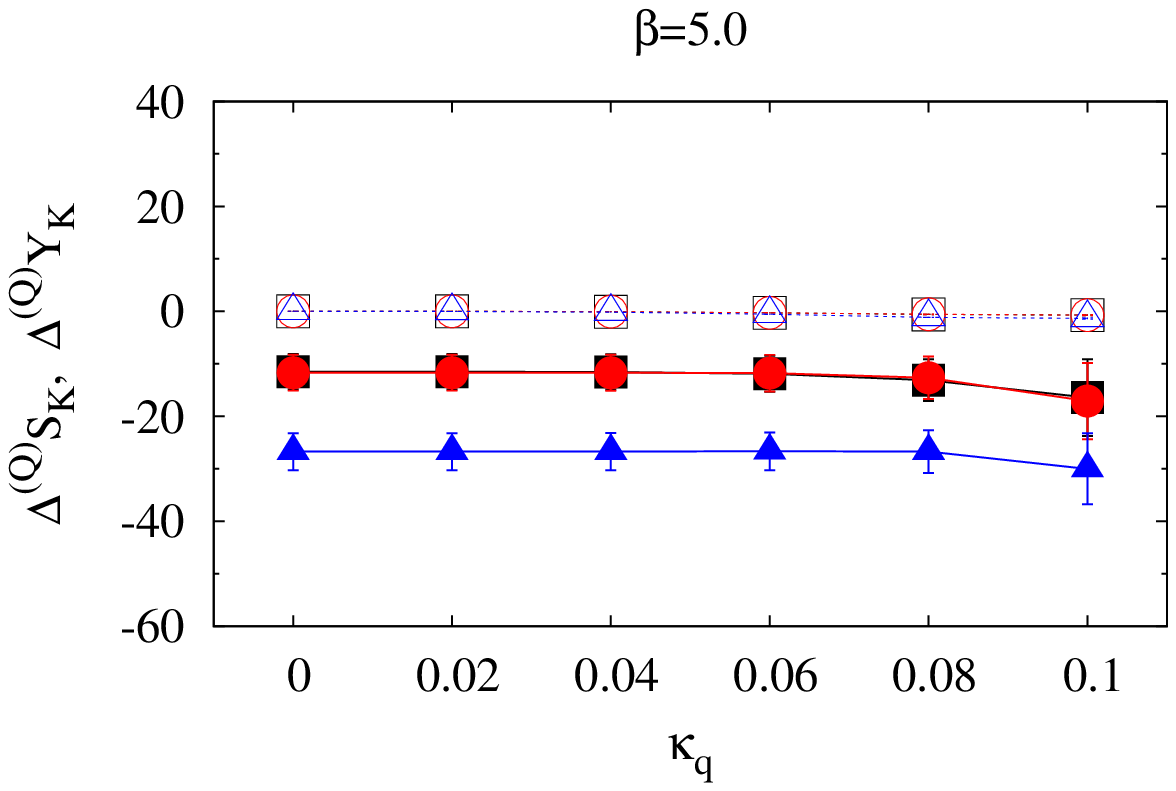} \\
  \hspace{-5ex}
  \includegraphics*[width=0.52 \textwidth,clip=true]
  {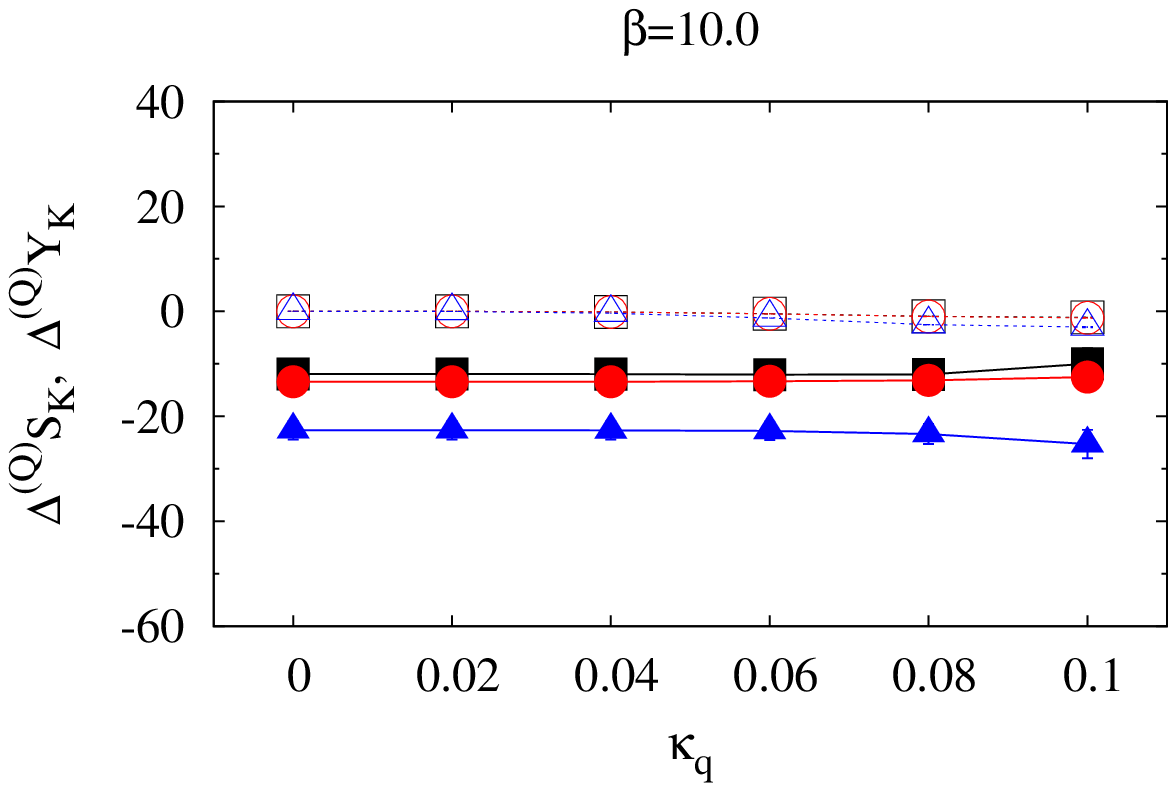} &
  \end{tabular}
\vspace{-2ex}
 \caption{$\Delta^{(Q)}S_K$ (filled) and $\Delta^{(Q)}Y_K$ (open) as a
  function of $\kq$ obtained on $16^3\times 4$.}
 \label{fig:16x4-7}
 \end{center}
\end{figure}

We define the following quantities,
\begin{eqnarray}
    \Delta^{(Q)}S_K
&=&     - \frac{\beta\,\beta_g}{6}
      \bigg(
          \frac{\langle S_g\, e^{N_f\kq^4 Y}\rangle_{\beta}^{(1)}}
               {\langle e^{N_f\kq^4 Y}\rangle_{\beta}^{(1)}}
        - \frac{\langle S_g\, e^{N_f\kq^4 Y}\rangle_{\beta}^{(0)}}
            {\langle  e^{N_f\kq^4 Y} \rangle_{\beta}^{(0)}}
      \bigg)
\ ,\\
    \Delta^{(Q)}Y_K
&=&   4\,N_f\,\kq^4\,\left(1 -8\kq\right)\,
      \bigg(
         \frac{\langle\, Y\,e^{N_f\,\kq^4\, Y}
	       \rangle^{(1)}_{\beta}}
              {\langle\, e^{N_f\,\kq^4\, Y}
	       \rangle^{(1)}_{\beta}}
       - \frac{\langle\, Y\,e^{N_f\,\kq^4\, Y}
               \rangle^{(0)}_{\beta}}
	      {\langle\, e^{N_f\,\kq^4\, Y}
	      \rangle^{(0)}_{\beta}}
     \bigg)
\ ,
\end{eqnarray}
where we omit the term including the anomalous dimension of quark mass.
Figure~\ref{fig:16x4-7} shows the $\kq$ dependence of the above two
quantities with $N_f=2$.
It turns out that the contribution of $\Delta^{(Q)}Y_K$ is much smaller
than that of $\Delta^{(Q)}S_K$ except for $\beta=100$.
Thus, omitting the term of the anomalous dimension does affect the final
result by much.

\begin{figure}[h]
 \begin{center}
  \begin{tabular}{cc}
  \hspace{-5ex}
  \includegraphics*[width=0.5 \textwidth,clip=true]
  {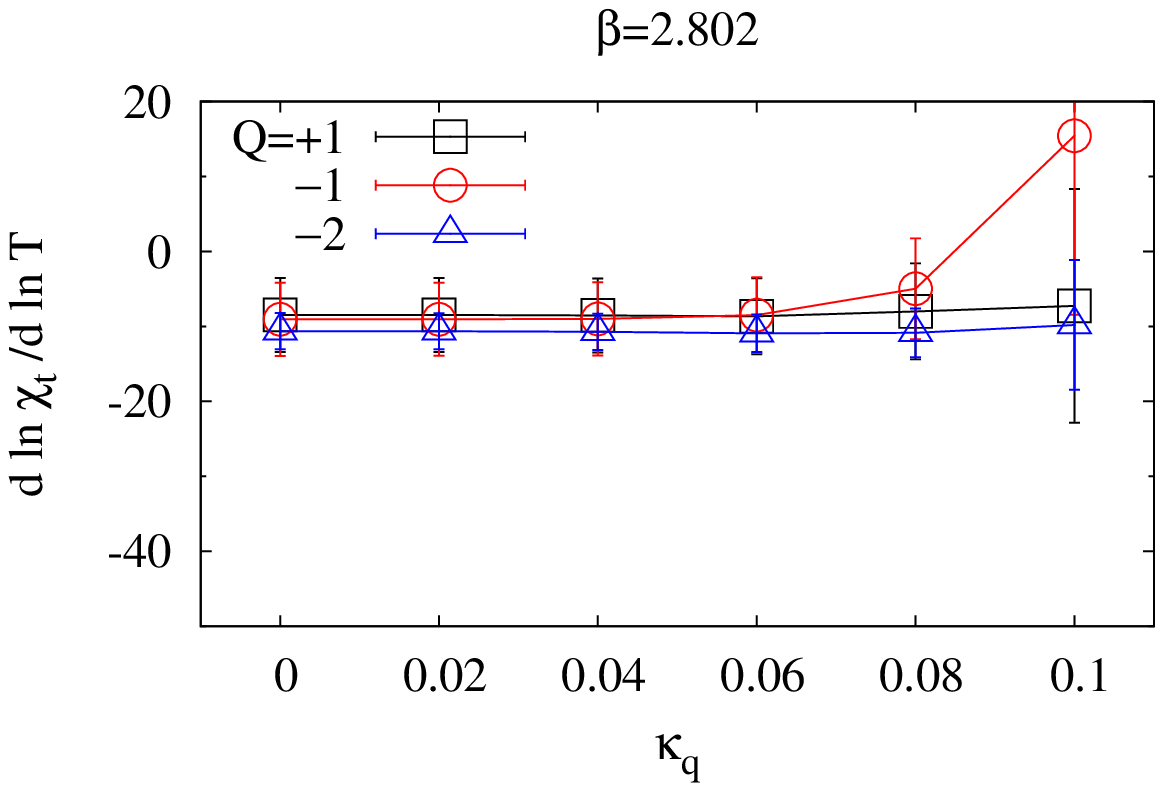}&
  \hspace{-5ex}
  \includegraphics*[width=0.5 \textwidth,clip=true]
  {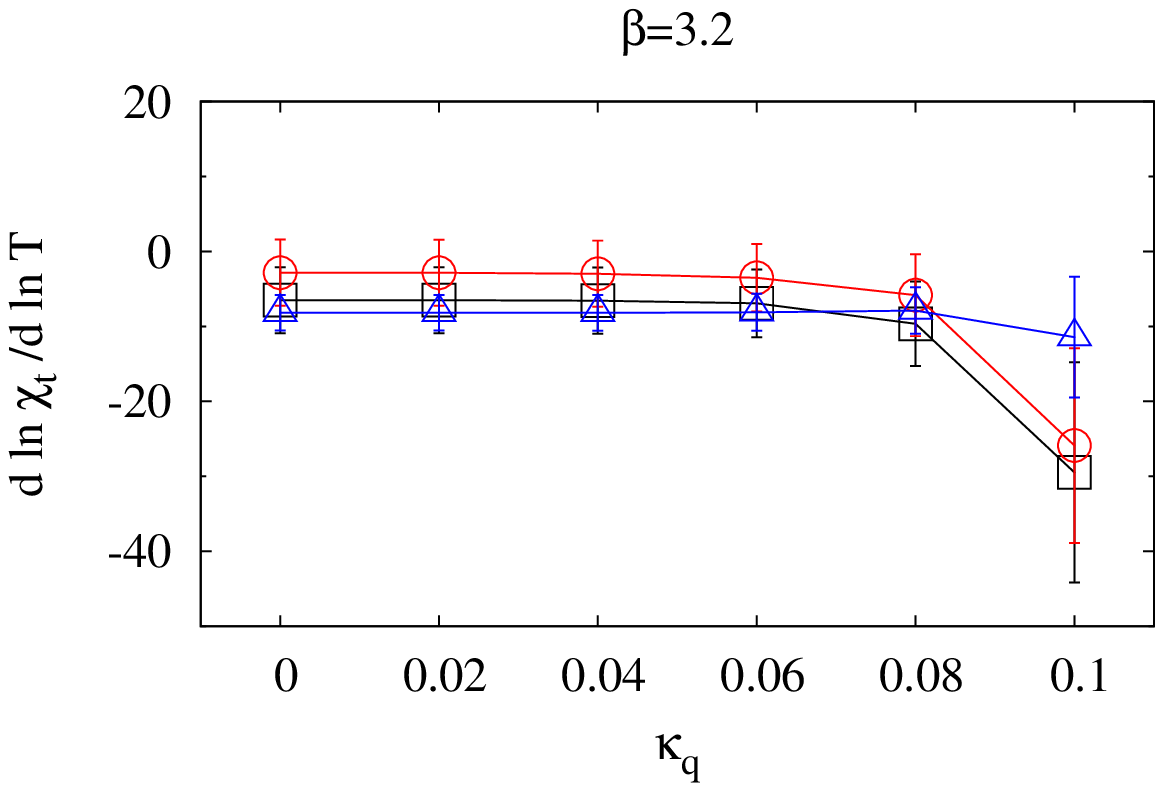} \\
  \hspace{-5ex}
  \includegraphics*[width=0.5\textwidth,clip=true]
  {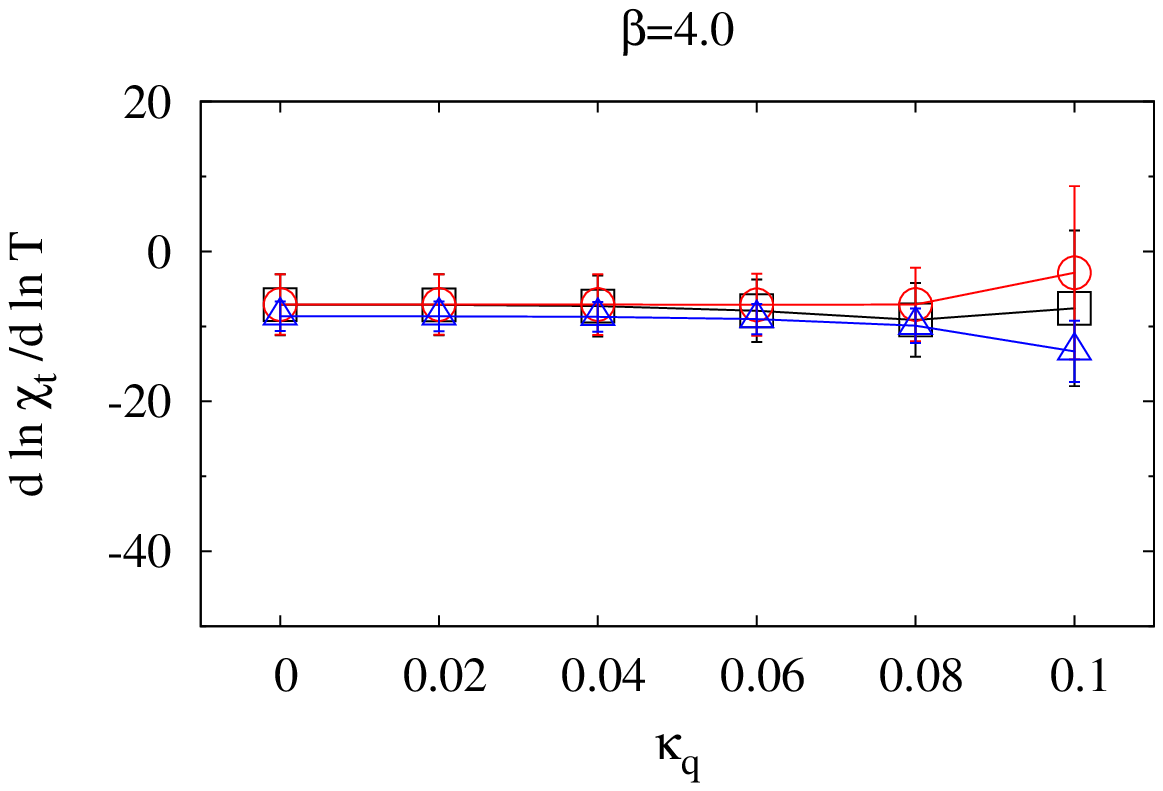} &
  \hspace{-5ex}
  \includegraphics*[width=0.5 \textwidth,clip=true]
  {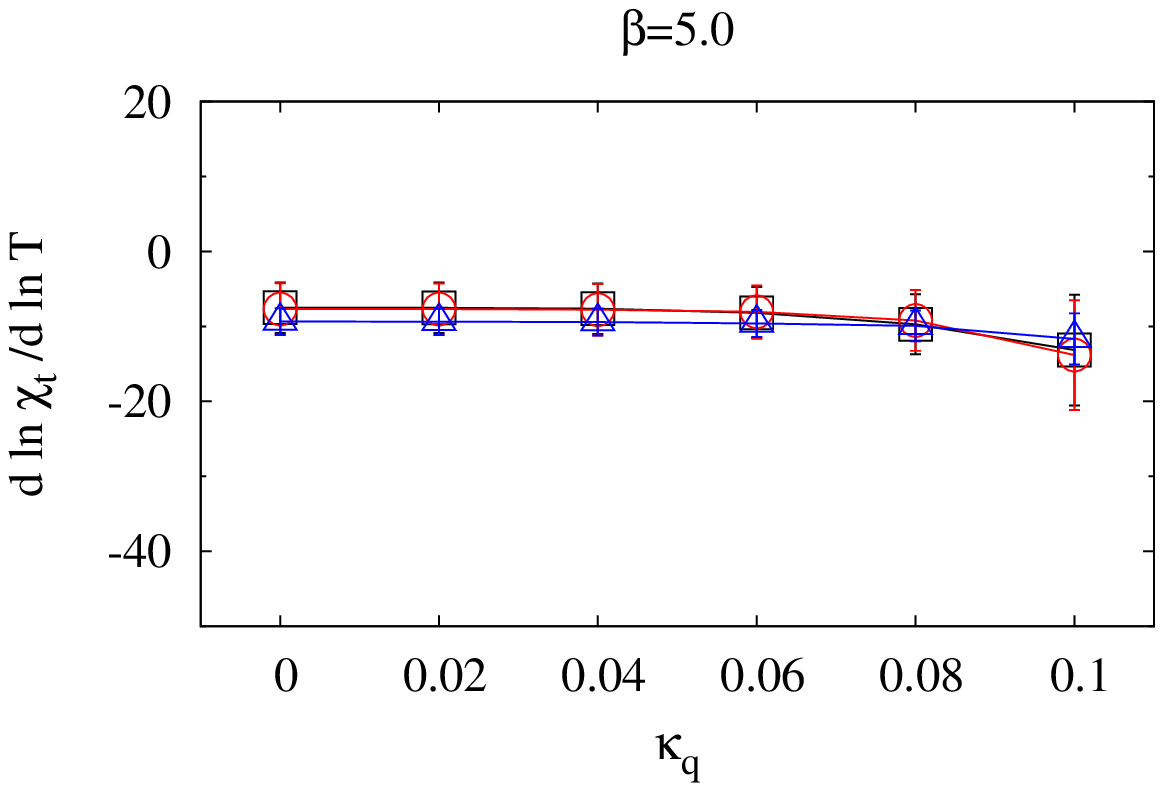} \\
  \hspace{-5ex}
  \includegraphics*[width=0.5 \textwidth,clip=true]
  {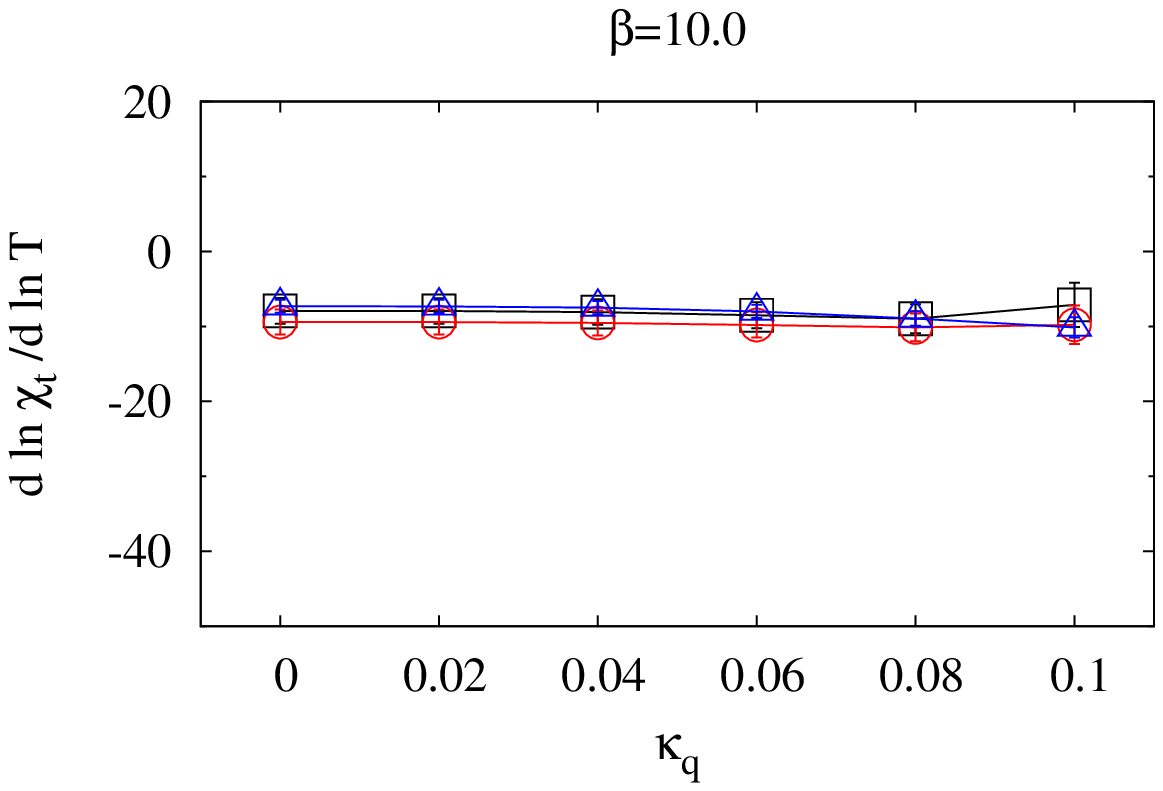} &
  \end{tabular}
\vspace{-2ex}
 \caption{$\kq$ dependence of $d\ln\chi_t/d\ln T$ from the HPE.}
 \label{fig:hpe-all}
 \end{center}
\end{figure}

$\pd\ln\chi_t/\pd\ln T$ is plotted in Fig.~\ref{fig:hpe-all}, which
shows no clear $\kq$ dependence up to $\kq=0.1$ except at $\beta=100$.
At $\beta=100$, the critical kappa is $\sim 0.125$, thus the result at
$\beta=100$ explores the relatively light quark mass region and may
indicate a tendency that $d\ln\chi_t/d\ln T$ decreases towards the
chiral limit, although the convergence of the HPE has to be checked.

\end{document}